\documentclass[11pt,a4paper]{emulateapj}
\usepackage{epsfig}
\usepackage{amsmath}
\usepackage{natbib}
\usepackage{verbatim}

\shorttitle{The saturation of SASI}

\begin{document}

\title{The saturation of SASI by parasitic instabilities}
\author{J\'er\^ome Guilet,
 Jun'ichi Sato  \&
 Thierry foglizzo}
\affil{Laboratoire AIM, CEA/DSM-CNRS-Universit\'e Paris Diderot, IRFU/Service d'Astrophysique, \\
CEA-Saclay F-91191 Gif-sur-Yvette, France. }

\begin{abstract}
The Standing Accretion Shock Instability (SASI) is commonly believed to be responsible for large amplitude dipolar oscillations of the stalled shock during core collapse, potentially leading to an asymmetric supernovae explosion. The degree of asymmetry depends on the amplitude of SASI, but the nonlinear saturation mechanism has never been elucidated. We investigate the role of parasitic instabilities as a possible cause of nonlinear SASI saturation.  As the shock oscillations create both vorticity and entropy gradients, we show that both Kelvin-Helmholtz and Rayleigh-Taylor types of instabilities are able to grow on a SASI mode if its amplitude is large enough. We obtain simple estimates of their growth rates, taking into account the effects of advection and entropy stratification. In the context of the advective-acoustic cycle, we use numerical simulations to demonstrate how the acoustic feedback can be decreased if a parasitic instability distorts the advected structure. The amplitude of the shock deformation is estimated analytically in this scenario. When applied to the set up of \cite{fernandez09a}, this saturation mechanism is able to explain the dramatic decrease of the SASI power when both the nuclear dissociation energy and the cooling rate are varied. Our results open new perspectives for anticipating the effect, on the SASI amplitude, of the physical ingredients involved in the modeling of the collapsing star.
\end{abstract}


\section{Introduction}
Despite decades of active research \citep{colgate66,bethe85}, the core collapse supernovae mechanism remains elusive. The failure of the most sophisticated 1D models to explode the majority of massive progenitors \citep{liebendorfer01} suggests that multidimensional effects are essential for a successful explosion. Understanding the hydrodynamical instabilities responsible for this symmetry breaking, and more specifically their nonlinear dynamics, is therefore required to understand the explosion mechanism. 

The region between the neutrinosphere and the shock deserves particular attention because two instabilities take place there: neutrino-driven convection \citep{herant92,herant94,burrows95,janka96,foglizzo06}, and the newly discovered Standing Accretion Shock Instability (SASI) \citep{blondin03, ohnishi06, foglizzo07, scheck08}. 2D simulations suggest that the complex fluid motions triggered by these instabilities could lead to a successful explosion either by helping the classical neutrino-driven mechanism \citep{buras06b, marek09, murphy08b} or by a new mechanism based on the emission of acoustic waves from the proto-neutron star (\cite{burrows06, burrows07a}, see however \cite{weinberg08}). The large scale ($l=1-2$) induced asymmetry could also explain the high kick velocities of newly formed neutron stars \citep{scheck04, scheck06} and may affect significantly their spin \citep{blondin07a, yamasaki08}.

The linear phase of the two instabilities has been described in details by \cite{foglizzo06, blondin06, foglizzo07,yamasaki07,fernandez09a}. Neutrino-driven convective modes with a large angular scale can be stabilized by a fast advection of matter through the gain region, whereas SASI is always dominated by large scale modes. This linear argument favors SASI as the cause of the prominent $ l=1-2 $ shock oscillations observed in the simulations. However, a theoretical understanding of the nonlinear development and saturation of SASI is still missing. This was highlighted by the unexpected dramatic decrease of the SASI power observed by \cite{fernandez09a} when both the nuclear dissociation energy and the cooling rate are varied. To shed more light on this issue, this paper proposes a predictive saturation mechanism for SASI. This is a first step toward understanding how the amplitude of SASI depends on the physical ingredients of the model (nuclear dissociation, equation of state, heating rate, rotation, magnetic fields).

We propose that the saturation takes place when a parasitic instability (also called secondary instability) grows on the dominant SASI mode. These parasites feed upon its energy and destroy its coherence, leading to a turbulent flow and the saturation of the growing SASI mode. This saturation mechanism by parasitic instabilities is similar to the one proposed for the saturation of the MRI by \cite{Goodman94, Pessah09a}.

Two types of instabilities are considered as potential parasites for the SASI mode: the Kelvin-Helmholtz instability (hereafter KHi) grows near a maximum of vorticity, and the Rayleigh-Taylor instability (hereafter RTi) grows on negative entropy gradients. In either case the cause of the instability lies within the SASI mode, and therefore the growth rate of these parasites increases with the amplitude of the shock oscillations. The parasites are thus able to affect significantly the dynamics of the flow only if the amplitude of SASI oscillations exceeds a certain threshold. The main objective of this work is to estimate this threshold.

Understanding the saturation mechanism of SASI requires in principle some understanding of the mechanism underlying its linear growth. The saturation scenario we propose is not restricted to a particular mechanism for the growth of SASI, but it can be understood more precisely in the framework of the advective-acoustic cycle \citep{foglizzo02, blondin03, ohnishi06, foglizzo07, scheck08, fernandez09a, foglizzo09}. In this mechanism, a shock deformation creates an entropy-vorticity wave, whose downward advection generates an acoustic feedback. This acoustic wave further deforms the shock, thus closing the unstable cycle. If the coherence of the advected wave were broken by a parasitic instability before it created the acoustic feedback, this cycle would be stabilized and SASI saturated.  

In Sect.~2, we explain our method and approximations. Sections 3 and 4 are devoted to the linear study of the KHi and RTi in our setup. In Sect.~5 we study how the development of the parasites breaks the coherence of a SASI mode and decreases of the acoustic feedback. In Sect.~6 we apply our results to the setup of \cite{fernandez09a} and compare our estimates with the results of their simulations. Finally, our results are discussed in Sect.~7 and summarized in Sect.~8.

\section{Method}
\subsection{Estimating the growth rate of the secondary instabilities} 
We estimate the local growth rate of the parasitic instability by using a simplified description of the linear SASI mode that keeps only the features that are essential for the physics of the instability. 

A SASI mode has a complex structure of entropy, vorticity and pressure perturbations which can be computed by a linear analysis \citep{foglizzo07, yamasaki07}. We focus on the advected structure of the SASI mode, where the parasitic instabilities operate. As we seek a local description of the parasites, we assume a planar geometry. Denoting by $z$ the vertical direction, and $x$ the transverse direction, the structure of the advected entropy/vorticity wave (denoted as $S$, $w$ respectively) is approximated by a sinusoid with a vertical wave number $ K_{ z} = \omega/v_0 $, where $\omega$ is the SASI frequency and $v_0$ is the flow velocity:
\begin{eqnarray}
S(z,t) & = & S_{ \rm{0}}(z) +  \Delta S \times \cos{\left(\omega t - K_z z\right)} 
	\label{flow1} \\
w(z,t) & = & \Delta w \times \cos{\left(\omega t - K_z z\right)} , 
	\label{flow2}
\end{eqnarray}
where $S_0$ is the entropy profile of the radial flow, and $\Delta S$, $\Delta w$ are the amplitudes of entropy and vorticity perturbations associated with the SASI mode. The advected entropy/vorticity wave in a SASI mode is actually tilted with respect to the horizontal direction, but we neglect this tilt for the sake of simplicity ($K_{ z}$ is a factor $ \gtrsim 5 $ larger than the horizontal wave number $K_{ x} \sim \sqrt{l(l+1)}/r $ in a typical SASI mode \citep{foglizzo07}). In what follows we omit the index $z$ and simply note $K=\omega/v_{ \rm{0}}$ the wave number of advected SASI perturbations. The horizontal wave number of each parasitic instability is denoted as $k$.
The notation $A_{ \rm{0}}$ refers to a quantity ($A$) in the stationary flow, unperturbed by SASI. Its perturbation by the SASI mode is denoted as $ \Delta A $, and $\delta A$ refers to a parasitic perturbation.
The growth rates of the KHi and RTi are directly related to the amplitude of the SASI mode through its profile of vorticity and entropy gradient. Non adiabatic cooling is taken into account in the shape of SASI eigenfunctions, but is neglected in the dynamical evolution of the parasites. We choose the usual adiabatic index of a relativistic gas $ \gamma = 4/3 $.
 
We estimate the stability of the flow described by Eq.~(\ref{flow1}) and (\ref{flow2}) in three steps, first by neglecting the background entropy gradient and global advection $v_0$, second by assessing the effects of a uniform entropy gradient $\nabla S_{ \rm{0}}$, and third by taking into account advection.

In the first two steps, a standard linear mode analysis can be used because the flow is stationary. The equations determining the evolution of the perturbations (e.g. Appendix~A of \cite{foglizzo99}) are solved numerically. A simple fitting formula for the maximum growth rate is proposed, as a function of the SASI amplitude and the background stratification. 

The third step is related to the concept of global versus local instability \citep{huerre90}, but the gradients in the direction of advection preclude the use of standard analytical techniques. If the fluid is advected too fast, the instability may be able to grow in a lagrangian way but would actually decay at a fixed radius as the perturbations are advected away. We use numerical simulations to measure the propagation speed of the parasitic instability in the $z$-direction, by perturbing the SASI mode over a limited region. Adjusting the measured speeds with a physical but approximate description of the propagation, we obtain a simple analytical estimate of the growth rate $\sigma_{ \rm{parasite}}$ of each parasitic instability, taking into account the SASI amplitude, the background stratification, and the advection speed (Eqs.~\ref{eqwiKH} and \ref{eqwiRT}). 

\subsection{Estimating the saturation amplitude}

The growth rate $\sigma_{ \rm{parasite}}$ is an increasing function of the SASI amplitude. A parasitic instability can affect the dynamics of SASI if its amplitude $ \delta A$ becomes comparable to the SASI amplitude $ \Delta A $. The growth of the ratio $ \delta A / \Delta A  $ requires $ \sigma_{ \rm{parasite}} > \sigma_{ \rm{sasi}} $. We use this criterion to estimate the saturation amplitude of SASI. \cite{Pessah09a} use a similar criterion to estimate the saturation amplitude of the MRI.

The criterion $\sigma_{ \rm{parasite}} = \sigma_{ \rm{sasi}} $ defines the minimum amplitude $ \Delta A_{\rm min}(r) $ of SASI  above which parasites can compete with SASI at a given radius $r$, despite advection and cooling. The parasitic instabilities can alter the growth of SASI only if their growth takes place in a region which is vital to the mechanism of SASI. For example, if the mechanism of SASI is interpreted as an advective-acoustic cycle, this cycle is most sensitive to the region between the shock and the deceleration region where most of the acoustic feedback is produced. 
Fortunately, as will be shown in Sect.~6, the local saturation amplitude $ \Delta A_{\rm min}(r) $ displays a broad minimum around the radius $ (r_{ \rm{*}}+r_{ \rm{sh}})/2 $, which defines a global saturation amplitude $ \Delta A_{\rm min} $ without much sensitivity on the details of the SASI mechanism.

\subsection{Limitations }
By focussing on the growth of parasitic instabilities, we ignore other nonlinear processes which could play a role in saturating the amplitude of SASI. Among them, the steepening of acoustic waves into shocks (e.g. \cite{fernandez09b}), the decoherence of the mode due to the finite displacement or velocity of the shock, or the exchange of energy by resonant mode coupling, could be important in some parameter range.

When compared to published simulations (Sect.~6), our estimate of the saturation amplitude of SASI based on parasitic instabilities is encouraging in view of the many simplifications inherent to our method:
our description of the parasitic growth neglects the spherical geometry of the flow, neglects the tilt of the SASI wave with respect to the horizontal direction, and assumes an adiabatic evolution of the parasites.
We further assume that SASI is dominated by a {\it single} SASI mode of finite amplitude, which we describe using a linear approximation. Steepened advected waves, induced by steepened acoustic waves reaching the shock, could affect the growth of parasitic instabilities by introducing larger vorticity and sharper entropy gradients.
We also neglect the production of entropy by acoustic waves steepening into shocks, and the creation of vorticity by the baroclinic interaction of entropy gradients with pressure waves.

\section{The Kelvin-Helmholtz instability (KHi)}

\subsection{The KHi in a sinusoidal velocity profile}

The KHi feeds on the kinetic energy available in shear flows. A necessary condition for its growth is the presence of a maximum in the absolute value of vorticity \citep{drazin81}. The linear growth of SASI creates a sinusoidal velocity profile 
\begin{equation}
v_{ x}\left(z \right) = \Delta v \sin{\left(K z \right) },
	\label{vprofile}
\end{equation}
with two such maxima per wavelength. \cite{heyvaerts83} have demonstrated that  this profile is indeed unstable to perturbations with a small horizontal wave number $k<K$, with a growth rate $ \sigma < k \Delta v $.

\begin{figure}
\centering
\includegraphics[width=\columnwidth]{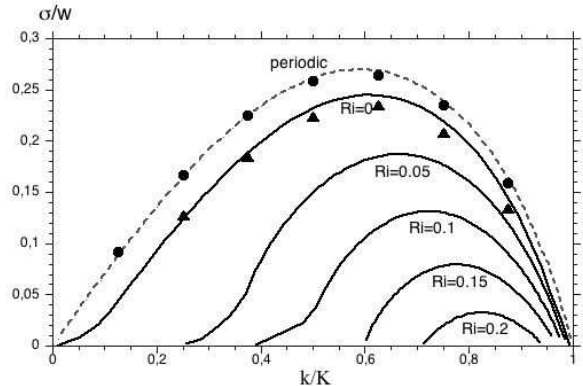}
 \caption{Growth rate $\sigma$ of the KHi as a function of the transverse wave number $k$. The dashed line is the eigenvalue for a sinusoidal velocity profile and periodic boundary conditions. The full lines show the effect of stratification with the Richardson number  Ri ranging from 0 to 0.2. The triangles and circles are measured from simulations where perturbations are localized (triangles) or extended (circles). $\sigma$ is normalized by the maximum vorticity $w$, and $k$ is normalized by the vertical wavenumber $K$ of the SASI mode.}
             \label{wiKH}%
\end{figure}

In order to estimate the maximum growth rate and the corresponding wavelength, we complement their analytical study by a numerical mode calculation where the flow profile is described by Eq.~(\ref{vprofile}), with periodic boundary conditions in $z=0$ and $z=1$. The results (Fig.~\ref{wiKH}) show good qualitative agreement with the expectation from \cite{heyvaerts83}: the maximum growth rate $ \sigma = 0.27K\Delta v$ is reached for a wave number $k=0.58K$, and the KHi is stable for $k>K$. The maximum KHi growth rate is thus a fraction of the maximum vorticity in the flow (see also \cite{foglizzo99}).
The Mach number ${\cal M}$ has little effect on the instability unless it approaches unity (we used $ {\cal M}=0.2$ in our mode calculation, with results almost identical to the incompressible case). 

\subsection{Effect of stratification on the KHi\label{stratKH}}

The buoyancy force in a stably stratified atmosphere is able to stabilize the KHi if $ {\rm Ri}  > 1/4 $, where the Richardson number ${\rm Ri} $ characterizes the relative strengths of buoyancy and shear (e.g. \cite{chandrasekhar61}):
\begin{equation}
{\rm Ri} \equiv \frac{N^{2}}{w^{2}},
\end{equation}
where $w \equiv \vec{\nabla} \times \vec{v}$ is the vorticity. In the absence of a composition gradient, the Brunt-V\"ais\"al\"a frequency $N$ is defined by:
\begin{equation}
N^{2} \equiv - \frac{\gamma-1}{\gamma}g\nabla S  .
	\label{brunt_vaisala}
\end{equation}
The entropy $S$ is here measured in dimensionless units $S\equiv \log(P/\rho^\gamma)/(\gamma-1)$.

In order to characterize this stabilization quantitatively in our specific geometry, we compute the eigenmodes of the KHi in a sinusoidal horizontal velocity field, embedded in a stable entropy gradient and a gravity field. The sinusoidal velocity profile, the background entropy gradient and gravity profiles have a constant amplitude in a limited region of space and are smoothly connected to zero outside this region. This region has a size of 3 vertical wavelengths. If far enough from the gradients, the choice of boundary condition (reflective or leaking condition) does not change the resulting mode. Results are shown in Fig.~\ref{wiKH}. The slight difference between the curve where the entropy gradient is zero ($ {\rm Ri} = 0$) and the periodic KHi is due to the limited size of the entropy gradient and gravity profiles. 

The maximum growth rate decreases linearly to zero as $ {\rm Ri} $ is increased, while the wave number at maximum varies from $ k \sim 0.6 K $ to $ k \sim 0.8 K $ (Fig.~\ref{wiKH}). The marginal stability is reached for a critical value of the Richardson number $ {\rm Ri} = 0.24  $ in close agreement with the expected value $ {\rm Ri} = 1/4  $.

\subsection{Effect of advection on the KHi\label{advKH}}

In the context of core collapse, the SASI vorticity wave is advected toward the neutron star. Using a physical argument, we first give a naive estimate of the speed at which an unstable perturbation can propagate against the stream, and evaluate the reduced growth rate in the presence of advection. Numerical simulations are used in order to obtain more accurate estimates.

\subsubsection{Physical argument}

The simplest illustration of the KHi (e.g. \cite{drazin81}) considers an incompressible fluid with a discontinuity of horizontal velocity: $v_{ x} = -\Delta v $ for $z<0$, and $v_{ x} = \Delta v $ for $z>0$. Unstable modes exist for any horizontal wave number with a growth rate $\sigma_{ 0} = k_{ x}\Delta v $, and their structure on both sides of the discontinuity is a decreasing exponential: $ \delta A \propto  e^{ -k_{ x}|z|} $. Viewed in a frame moving with a vertical velocity $ v_{ z} $, the time dependence of the perturbation is: $ \delta A \propto  e^{ \left(\sigma_{ 0}-k_{ x}v_{ z}\right)t} $ at a given height $z$ above the discontinuity. A condition for the perturbation to be growing at a given radius can be deduced: $ \sigma_0 > k_{ x}v_{ z} $, which may be interpreted as a propagation speed $v_{ \rm{prop}}$ of the KHi equal to:
\begin{equation}
v_{ \rm{prop}} = \frac{\sigma_0}{k_{ x}}.
	\label{propagation}
\end{equation}
The local growth rate in the frame moving with respect to the discontinuity is thus decreased as follows: 
\begin{equation}
\sigma = \sigma_{ 0}\left(1 - \frac{v_{ z}}{v_{ \rm{prop}}} \right).
	\label{growth}
\end{equation}

Of course this physical argument is too simple to be directly applicable to the case of a sinusoidal shear wave advected downward, where each unstable shear layer is followed by another one. Could the propagation be faster if these adjacent shear layers cooperate ?

\subsubsection{Numerical simulations of the KHi}
\begin{figure}
\centering
\includegraphics[width=\columnwidth]{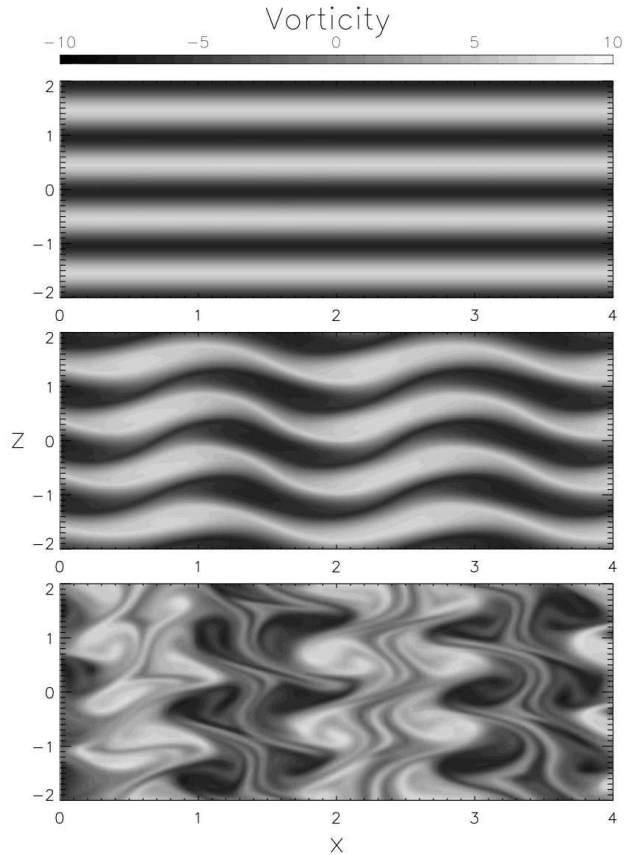}
 \caption{Different stages in the evolution of the KHi on a sinusoidal velocity profile. The \emph{upper plot} is the initial condition: a sinusoidal transverse velocity wave with random perturbations localized between $-0.5 <z< 0.5$ (one wavelength of the stationary flow). The \emph{middle plot} shows the time when the KHi just reached a nonlinear amplitude, the mode structure is still clear. Finally the \emph{bottom plot} shows a more developed nonlinear stage of the instability. In all plots, the grayscale represents the vorticity.}
             \label{imageKH}%
\end{figure}

Using the code RAMSES \citep{Teyssier02, Fromang06} we performed numerical simulations of a sinusoidal velocity profile described by Eq.~(\ref{vprofile}) (with $K=2\pi$, $ \Delta v = 1 $, $c=5$). The computational domain was a box $-8<z<8$, $ 0< x<8$, with periodic boundary conditions. Perturbations of the vertical velocity, localized between $ z= -0.5 $ and $ z= 0.5 $ (one wavelength), have been added to the stationary flow, with an amplitude of  $ 10^{-4} \Delta v $. The runs presented here have a resolution of $ 1024 \times 512 $, but we checked that the results do not depend on the resolution.

The different phases of the KHi growth are illustrated by three snapshots in Fig.~\ref{imageKH}. The wavelength $ \sim 1.8 $ of the dominant feature agrees with the expected wavelength of the fastest mode in Fig.~\ref{wiKH} ($k/K\sim 0.55$). As a consistency check, the growth rates have been measured for different wave numbers and compared successfully with the linear analysis in Fig.~\ref{wiKH}\footnote{The growth rates (triangles) are slightly smaller than the linear values with periodic boundary condition, but are close to those of the modes whose spatial extent is restricted to three wavelengths (Sect.~3.2). Note that they are not expected to match exactly, but the fact that they have similar values suggests a common physical origin of the reduced growth rate, namely the restricted extent of the region where the KHi grows. For comparison we also ran a simulation where the whole flow was perturbed instead of the region $ z = [-0.5, 0.5] $. The growth rates (circles) are in very good agreement with the modes computed using periodic boundary conditions (dashed line). }.

\begin{figure}
\centering
\includegraphics[width=\columnwidth]{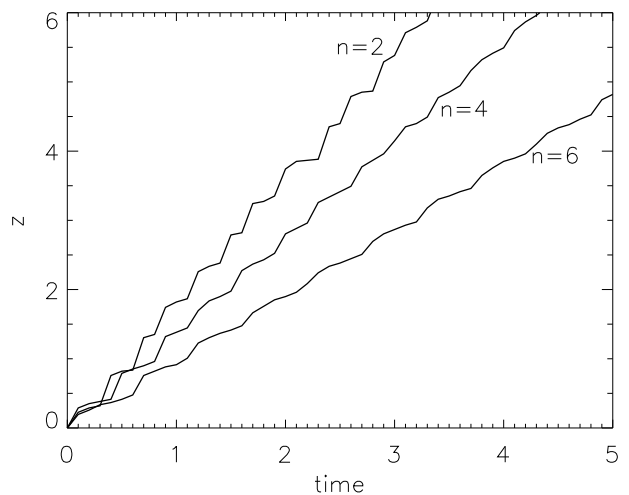}
 \caption{Propagation of the KHi along the $z$ direction. The three curves represent different modes with a number of horizontal wavelengths ranging from 2 (upper curve) to 6 (lower curve). The plotted quantity, $z_{ \rm +}\left(t\right)$, corresponds to the spatial extent in which the KHi has reached an amplitude of $\delta v = 10^{-4}\Delta v$. The curves were shifted in time so that they start to deviate from 0 at $ t=0$.}
             \label{propagationKH}%
\end{figure}

In the two bottom plots in Fig.~\ref{imageKH} the KHi mode has already propagated downward and upward from the initial perturbations.  For a quantitative study of the propagation speed, we measured the spatial range $ \left[-z_{ \rm +},z_{ \rm +}\right]$ in which the KHi has reached an arbitrary amplitude (say $\delta v =10^{-4}\Delta v$), as a function of time for different wave numbers. As expected from 3.3.1, the large wavelength perturbations propagate faster than the short wavelength ones (Fig.~\ref{propagationKH}). 

\begin{figure}
\centering
\includegraphics[width=\columnwidth]{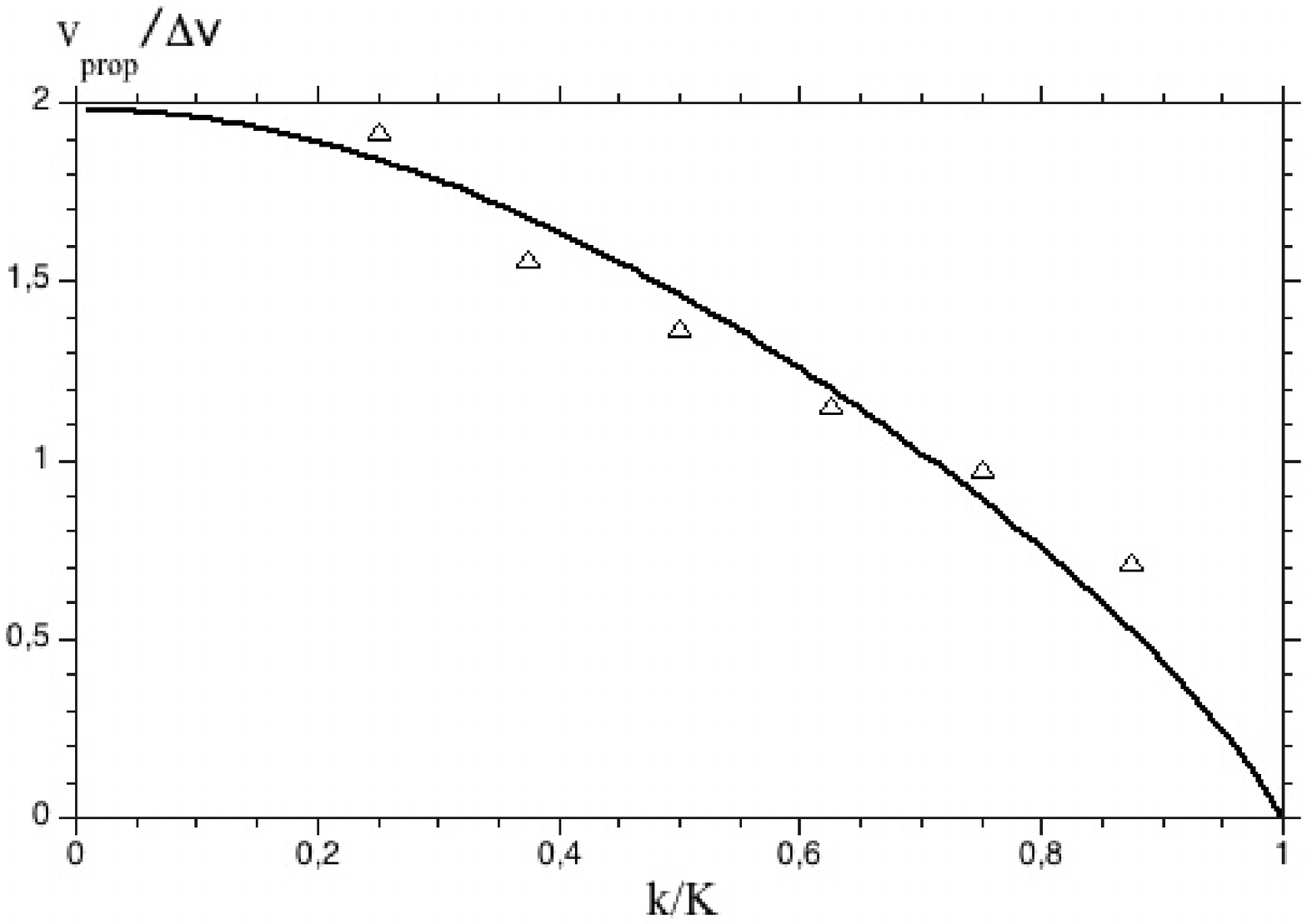}
\includegraphics[width=\columnwidth]{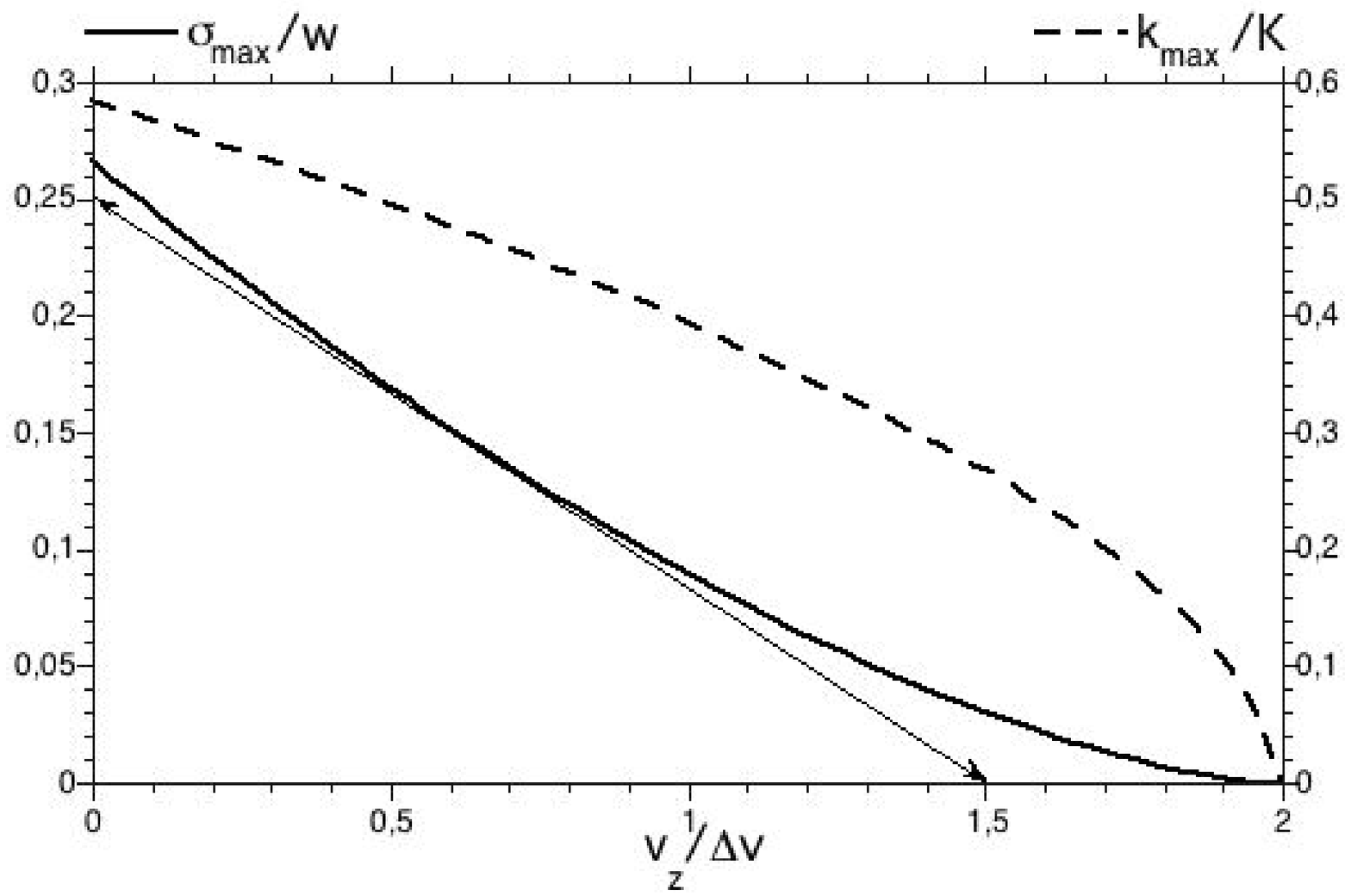}
 \caption{\emph{Upper plot:} Propagation speed $v_{\rm prop}$ of the KHi as a function of the wave number $k$, measured in the numerical simulations (triangles). The full line shows the quantity $ 2.8 \sigma/k $, where $\sigma$ is the growth rate computed in the linear analysis of Sect.~3.1. The factor $ 2.8$ has been adjusted in order to match the results of the simulations. \emph{Bottom plot}: maximum growth rate (full line, left axis) and associated wave number (dashed line, right axis) as a function of the advection velocity. The thin straight line with arrows is a linear approximation of the growth rate (Eq.~(\ref{sigkhmax})) .}
             \label{advectionKH}%
\end{figure}

The propagation velocity measured in the simulations is typically a factor $ 2.8$ higher than the quantity $ \sigma/k $ suggested in 3.3.1, but the dependence on $k$ is similar (upper plot in Fig.~\ref{advectionKH}). We interpret this high propagation speed as the sign that adjacent shear layers do cooperate.

Approximating the propagation speed by $v_{ \rm{prop}} = 2.8\times \sigma/k $, the wavelength dependence of the growth rate for a given advection velocity is estimated using Eq.~(\ref{growth}). The maximum growth rate is smaller, and obtained for a longer wavelength than without advection, because shorter wavelengths propagate more slowly (Fig.~\ref{advectionKH}). The maximum growth rate $ \sigma_{ \rm{KHmax}} $ decreases almost linearly with the advection velocity $v_{z} $. We approximate this curve linearly as follows: 
\begin{equation}
\sigma_{ \rm{KHmax}} = \sigma_{ \rm{0KH}}\left(1 - {v_{z}\over v_{ \rm{eff0}}} \right),\label{sigkhmax}
\end{equation}
where $ \sigma_{ \rm 0KH} = 0.25\Delta w $ is the maximum KHi growth rate without advection, and $v_{ \rm eff0} $ is an effective propagation speed which we estimate to be (in the absence of stratification) $v_{ \rm eff0} \sim 1.5\times \Delta v \simeq 1.5 \Delta w/K$. 
Note that $v_{ \rm prop} $ in Eq.~(\ref{growth}) depends on the wavelength $k$ of the perturbation, whereas $v_{ \rm eff} $ in Eq.~(\ref{sigkhmax}) is independent of $k$, because $\sigma_{ \rm{KHmax}}$ is the growth rate maximized over all wavelengths. 
 $v_{ \rm eff} $ can be interpreted as an average propagation speed of the modes growing fastest at different advection speeds.

\subsection{Analytical estimate of the KHi growth rate with both advection and stratification}

In the presence of stratification, Eq.~(\ref{sigkhmax}) becomes :
\begin{equation}
\sigma_{ \rm{KH}} = \sigma_{ \rm{KHstrat}}\left(1 - {v_{z}\over v_{ \rm{eff}}} \right),\label{sigkhmaxbis}
\end{equation}
where $\sigma_{ \rm{KHstrat}} = \sigma_{ \rm{0KH}}\left(1 - {\rm Ri}/{\rm Ri_{0}} \right) $ is the maximum growth rate of the KHi in the presence of stratification but in the absence of advection, and $v_{ \rm{eff}}$ is the effective propagation velocity \emph{in the presence of stratification.} Eq.~(\ref{propagation}) suggests that the propagation speed is proportional to the growth rate, giving :
\begin{equation}
 v_{ \rm{eff}} = \frac{\sigma_{ \rm{KHstrat}}}{\sigma_{ \rm{0KH}}}v_{ \rm{eff0}},
 	\label{vpropstrat}
\end{equation}
Injecting Eq.~(\ref{vpropstrat}) into Eq.~(\ref{sigkhmaxbis}) then gives the following formula :

\begin{equation}
\sigma_{ \rm KH} = \sigma_{ \rm 0KH}\left(1 - \frac{ {\rm Ri} }{ {\rm Ri_{0}} } - \frac{v_{ z}}{v_{ \rm eff0}} \right), \label{sigkh1}
\end{equation}
where $\sigma_{\rm 0KH} = 0.25\Delta w $ is the growth rate in the absence of stabilizing effect, $ {\rm Ri_{0}} = 0.24 $, and $v_{\rm eff0} = 1.5 \Delta w/K $.
Equation~(\ref{sigkh1}) can be rewritten as a function of the SASI amplitude:
\begin{equation}
\sigma_{ \rm KH} = 0.25\Delta w - 1.04\frac{ N_{0}^{2} }{\Delta w } - \frac{ Kv_{z} }{6}.
	\label{eqwiKH}
\end{equation}
In a SASI eigenmode, the vorticity $ \Delta w $ at a given position in the flow is directly proportional to the relative shock displacement $\Delta r/r_{\rm sh}$. The factor $w_{ \rm sasi}(r)$ defined by $ \Delta w \equiv  w_{ \rm sasi} \Delta r/r_{\rm sh} $ depends on the radius, and is a result of the linear mode analysis of SASI.

A local saturation amplitude of SASI due to the parasitic growth of the KHi, at a given radius, is deduced from Eq.~(\ref{eqwiKH}) and the criterion $ \sigma_{\rm KH} = \sigma_{\rm sasi} $:
\begin{equation}
\frac{\Delta r}{r_{\rm sh}} = \frac{ \left\lbrack \left( {1\over3}Kv_{z}+2\sigma_{\rm sasi} \right)^{2} + 4.2N_{0}^{2} \right\rbrack^{1/2}+ {1\over3}Kv_{z}+2\sigma_{\rm sasi} }{w_{\rm sasi}}.
\label{satKH}
\end{equation}
The discussion of a global saturation amplitude is postponed to Sect.~6.

\section{The Rayleigh-Taylor instability (RTi)}

\subsection{Simple RTi in a sinusoidal entropy profile}

The RTi feeds on the potential energy available when a low entropy fluid is sitting on top of a higher entropy one, in a gravitational acceleration $g$. Buoyancy is characterized by the Brunt-V\"ais\"al\"a frequency $N$ (Eq.~(\ref{brunt_vaisala})). The flow is unstable if $ N^{2} $ is negative, and the typical growth rate $\sigma_{\rm RT}$ of short wavelengths perturbations is: 
\begin{equation}
\sigma_{\rm RT} \equiv \left(-N^{2}\right)^{1\over 2} = \left( \frac{\gamma-1}{\gamma}g\nabla S\right)^{1\over 2}.
\end{equation}
In the sinusoidal entropy profile created by SASI, the sign of the entropy gradient $\nabla S$ changes every half wavelength: the entropy wave is made of adjacent layers of stably stratified and Rayleigh-Taylor unstable fluid. The RTi is expected to grow fastest where the entropy gradient is most negative. 

\begin{figure}
\centering
\includegraphics[width=\columnwidth]{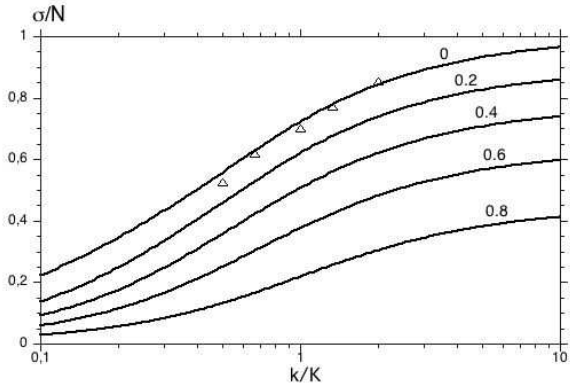}
 \caption{Growth rate of the RTi as a function of the transverse wave number. The full lines are the result of the linear analysis of a sinusoidal entropy profile embedded in a stable background entropy gradient and a gravity field. The different curves show different values of the background entropy gradient ranging from $\nabla S_{0} = 0$ (upper curve) to $\nabla S_{0} = 0.8\nabla\Delta S $ (lower curve). The triangles show the growth rate measured in the numerical simulations without background entropy gradient.}
             \label{wiRT}%
\end{figure}

By computing the eigenmodes of a sinusoidal entropy profile with a vertical wavenumber $K$ in a constant gravity field, we verified that the growth rate $\sigma\sim \sigma_{\rm RT}$  at short wavelength (Fig.~\ref{wiRT}), and $\sigma\sim 0.75\sigma_{\rm RT}$ if $ k \sim K$. In this calculation, as in Sect.~\ref{stratKH}, the entropy gradient and gravity profiles have a constant amplitude over a limited region of space (three vertical wavelengths), and are smoothly connected to zero outside this region. 

\subsection{Effect of stratification on the RTi\label{stratRT}}

Just as KHi, the RTi can be stabilized by the presence of a stable entropy gradient in the stationary flow $ \nabla S_{0} $. The flow is stable where the stationary gradient $ \nabla S_{0} $ is stronger than the SASI gradient $\nabla \Delta S$. The maximum growth rate is expected to scale as the Brunt-V\"ais\"al\"a frequency associated with the most negative entropy gradient:
\begin{equation}
\sigma_{\rm RT} =  \left\lbrack \frac{\gamma-1}{\gamma}g\nabla \left( \Delta S + S_{0} \right)\right\rbrack^{1\over2}.
	\label{wiRTstrat}
\end{equation}
This was checked by adding a background positive entropy gradient to the previous mode analysis. The resulting growth rate, shown in Fig.~\ref{wiRT}, follows Eq.~(\ref{wiRTstrat}) at short wavelengths. At longer wavelength, the RTi is stabilized slightly faster. The cause may be that these larger scale modes, in addition to grow on less intense negative entropy gradient, are also sensitive to the higher positive entropy gradient (which are stable).

\subsection{Effect of advection on the RTi\label{advRT}}

\subsubsection{Physical argument\label{physRT}}

The simple physical argument applied to the KHi in Sect.~3.3.1 can be adapted to the case of the RTi by considering an incompressible fluid with a discontinuity of density with $\rho= \rho_{0}-\Delta\rho $ for $z<0$ and  $\rho= \rho_{0}+\Delta\rho $ for $z>0$. Perturbations with a horizontal wavenumber $k_x$ are unstable with a growth rate $\sigma_{0} = (k_{x}g\Delta\rho/\rho )^{1/2} $ (e.g. \cite{chandrasekhar61}). The vertical structure of the RTi mode on both sides of the discontinuity is a decreasing exponential: $ \delta A \propto  e^{ -k_{x}|z|} $ due to the incompressible nature of the flow ($ k_{z}^{2} + k_{x}^{2} = 0 $). 
Using the same argument as in Sect.~\ref{advKH}, the estimated propagation speed is $ v_{\rm prop} = \sigma_{0}/k_{x} $.
In an entropy wave, each unstably stratified layer is followed by a stably stratified one. Unlike the instability of a vorticity wave, adjacent layers are not expected to cooperate. The vertical propagation of the RTi is expected to be accelerated by unstable layers and decelerated by stable ones. 

\subsubsection{Numerical simulations of the RTi}

\begin{figure}
\centering
\includegraphics[width=\columnwidth]{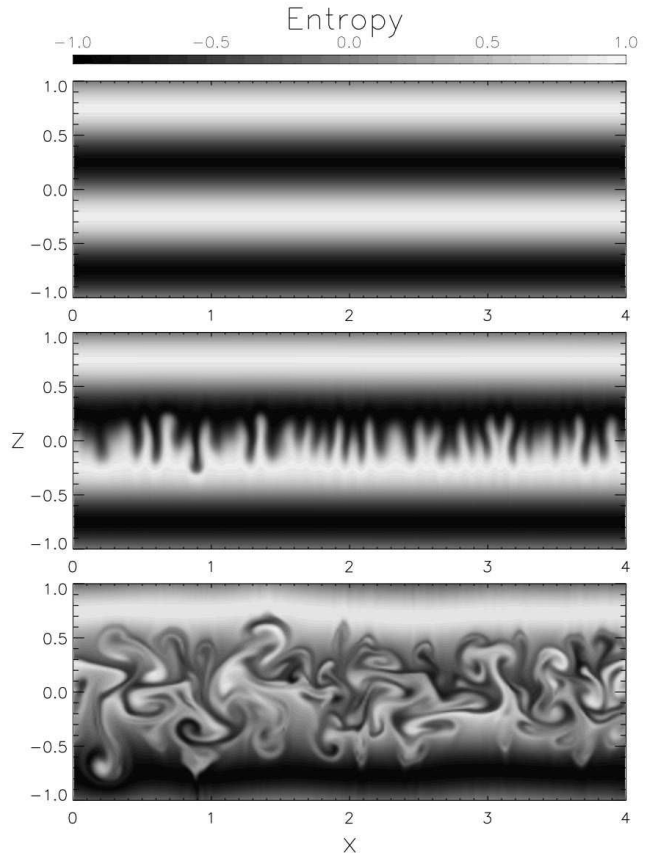}
 \caption{Different stages in the evolution of the RTi on a sinusoidal entropy profile. \emph{The upper plot} is the initial condition: a sinusoidal entropy wave embedded in a constant gravity field, with random perturbations of the vertical velocity localized between $-0.2 <z< 0.2$. \emph{The middle plot} shows the time when the RTi just reached a nonlinear amplitude, the mode structure is still clear. Finally \emph{the bottom plot} shows a more developed nonlinear stage of the instability. In all plots, the grayscale represents the entropy.}
             \label{imageRT}%
\end{figure}

In order to measure the propagation speed of the RTi, we performed numerical simulations of a sinusoidal entropy profile $S(z)$ described by:
\begin{equation}
S\left(z \right) = \Delta S \sin{\left(K z \right) }.
	\label{Sprofile}
\end{equation}
In units of the vertical wavelength $2\pi/K\equiv1$, we chose a computational domain $-8<z<8$ and $ 0< x<8$, with periodic boundary conditions in the $x$ direction, and a resolution of $ 1024 \times 512 $. The entropy oscillations described by Eq.~(\ref{Sprofile}) and the gravity profile have been restricted to $ -5<z<5 $, the medium being uniform for $ |z|>5 $. At $|z|=8$ we imposed a zero gradient boundary condition. The different phases of the RTi growth are illustrated by three snapshots in Fig.~\ref{imageRT}. The RTi grows at all scales resolved by the grid. Numerical convergence has been checked for the modes studied in the following ($k/K \gtrsim 0.5 $), and the growth rates agree well with the linear analysis (Fig.~\ref{wiRT}).

Perturbations of the vertical velocity localized in a narrow layer $  -0.2 < z < 0.2 $ have been added to the stationary flow with an amplitude of $ 0.01 \%$ of the sound speed $c_0$ in order to follow their propagation. 

\begin{figure}
\centering
\includegraphics[width=\columnwidth]{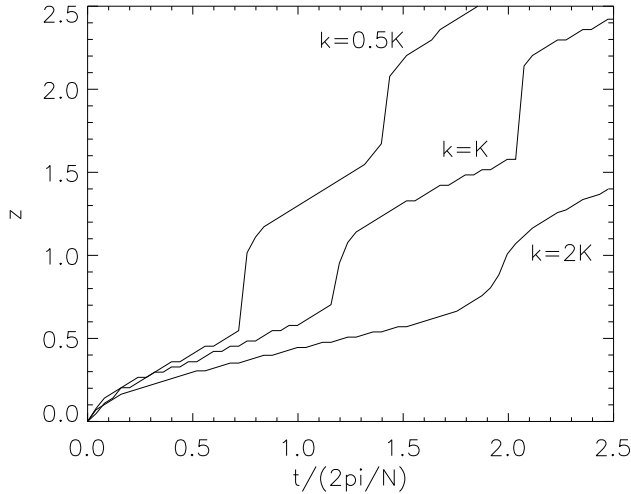}
 \caption{Propagation of the RTi along the $z$ direction. The three curves represent modes with a wave number ranging from $ k=0.5K $ (upper curve) to $k=2K$ (lower curve). The plotted quantity,  $z_{\rm +}\left(t\right)$, corresponds to the spatial extent in which the RTi has reached an amplitude of $\delta v = 10^{-4} c $. The curves have been shifted in time so that they start to deviate from 0 at $ t=0$.}
             \label{propagationRT}%
\end{figure}

\begin{figure}
\centering
\includegraphics[width=\columnwidth]{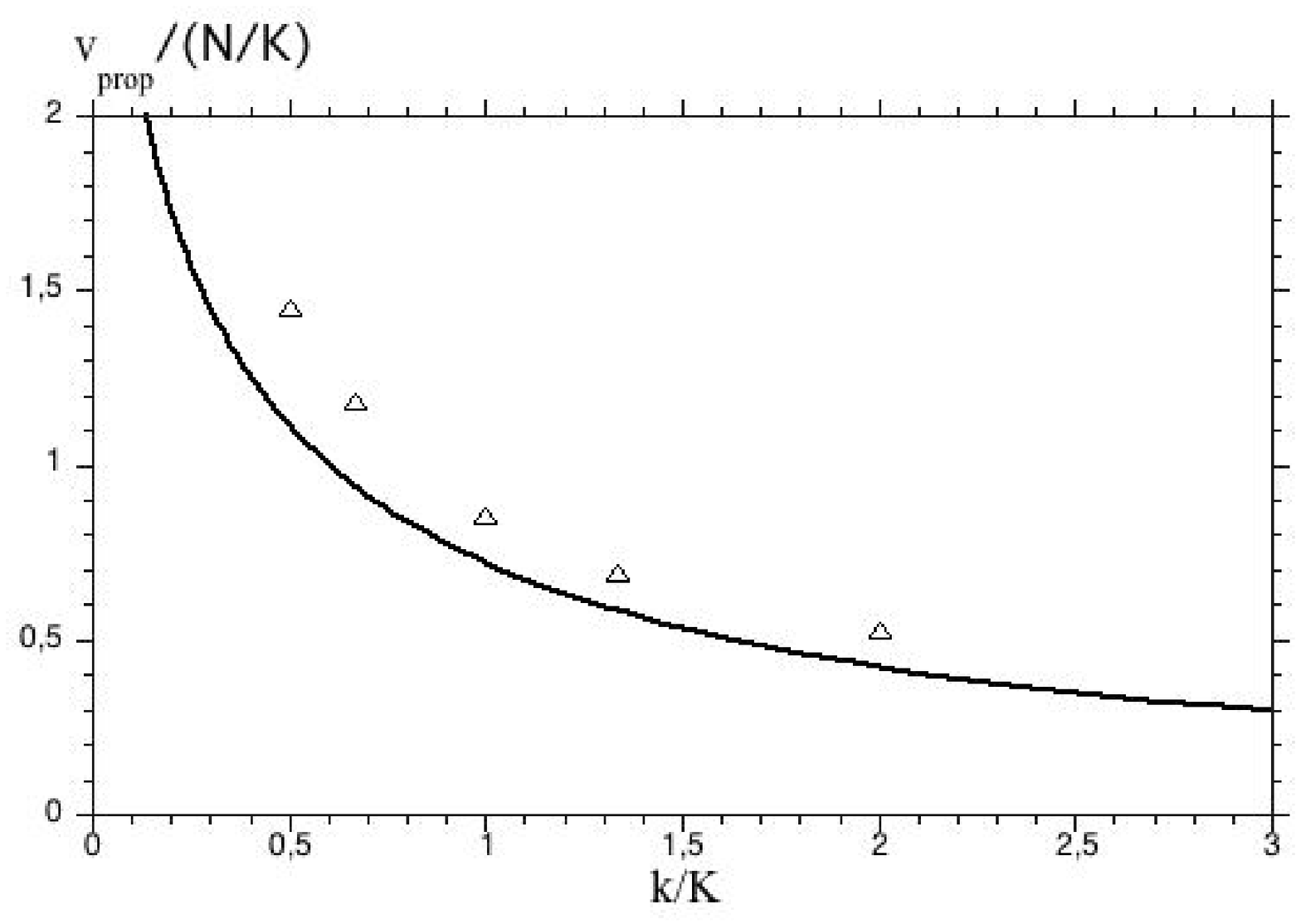}
\includegraphics[width=\columnwidth]{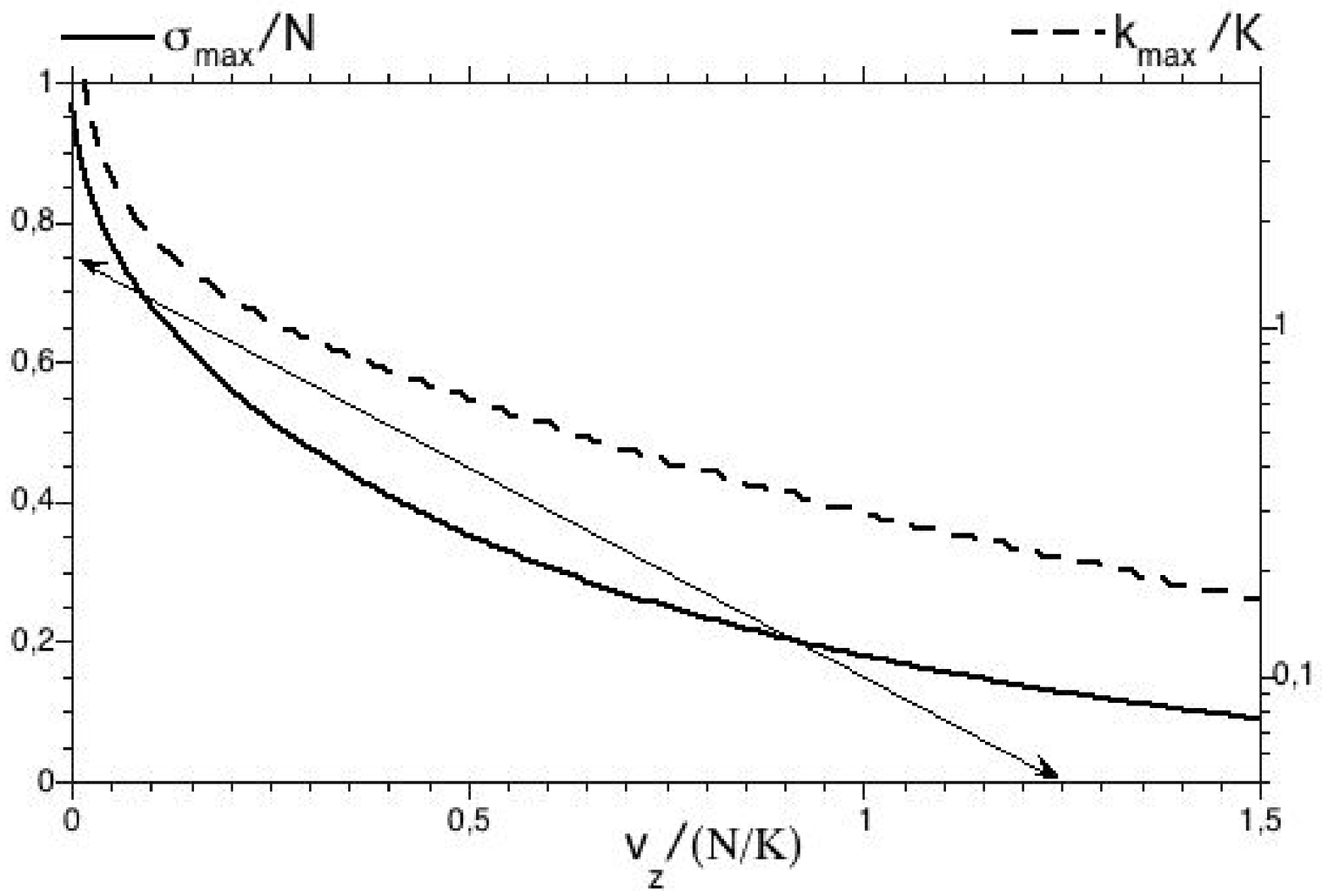}
 \caption{\emph{Upper plot:} propagation speed of the RTi as a function of its wave number $k$. The triangles are measured in the simulations. The black line is the quantity $ \sigma/k $ estimated from the linear mode calculation. \emph{Bottom plot:} maximum growth rate of the RTi as a function of the advection velocity (full thick line) and corresponding wave number (dashed line). The thin line with arrows illustrates the linear approximation of the growth rate used in Eq.~(\ref{lineaRT}).}
             \label{advectionRT}%
\end{figure}

As expected in the previous Section, the propagation is less regular than for the KHi: it is slow in the stably stratified regions and faster in the unstable ones (Fig.~\ref{propagationRT}). It is however possible to measure a global propagation speed, which is 
remarkably close to the estimate $ \sigma/k $ deduced from the mode analysis (upper plot in Fig.~\ref{advectionRT}). The propagation speed is well approximated by $v_{\rm prop} = 1.25\sigma/k $. 

Using Eq.~(\ref{growth}), the RTi growth rate for a given advection speed is expressed as a function of the wave number. It is maximum at a longer wavelength than in the absence of advection, because shorter wavelengths propagate more slowly. We approximate linearly the decrease of the maximum RTi growth rate $ \sigma_{\rm RTmax} $ when the advection velocity $v_{z} $ increases (bottom plot in Fig.~\ref{advectionRT}): 
\begin{equation}
\sigma_{\rm RTmax}= \sigma_{\rm 0RT}\left(1 - {v_{z}\over v_{\rm eff0} }\right), \label{lineaRT}
\end{equation}
where  $ \sigma_{\rm 0RT} = 0.75\Delta N $ is the RTi growth rate without advection and $v_{\rm eff0} $ is an effective propagation speed which we estimate to be $v_{\rm eff0} \simeq 1.25 \Delta N/K$.

\subsection{Analytical estimate of the RTi growth rate with both advection and stratification}
By a similar reasoning as in Sec.~3.4, the RTi growth in the presence of both advection and stratification is approximated using the results of Sects.~\ref{stratRT} and \ref{advRT}:
\begin{equation}
\sigma_{\rm RT} = 0.75\left\lbrack \frac{\gamma-1}{\gamma}g\nabla\left( \Delta S + S_{0} \right)\right\rbrack^{1\over 2} - 0.6Kv_{z}.
	\label{eqwiRT}
\end{equation}
The entropy gradient $ \nabla\left(\Delta S  \right) $ in a SASI mode is proportional to the relative shock displacement $\Delta r/r_{\rm sh}$. Defining $\nabla S_{\rm sasi}$ by $ \nabla\left(\Delta S  \right) \equiv\nabla S_{\rm sasi}\Delta r/r_{\rm sh}$, we use the criterion $ \sigma_{\rm RT} = \sigma_{\rm sasi} $ to obtain an explicit estimate of the local saturation amplitude of SASI due to the parasitic growth of the RTi:
\begin{equation}
\frac{\Delta r}{r_{\rm sh}} = \frac{\nabla S_{0}}{\nabla S_{\rm sasi}} + \frac{ \gamma }{0.56\left( \gamma -1\right)g\nabla S_{\rm sasi} }\left(\sigma_{\rm sasi} + 0.6k_{\rm sasi}v_{z} \right)^{2}.
	\label{satRT}
\end{equation}

\section{Acoustic feedback in the presence of parasitic instabilities}

The analytical estimates of the saturation amplitude obtained in Eq.~(\ref{satKH}) and (\ref{satRT}) from the criterion $\sigma_{\rm parasite} = \sigma_{\rm sasi} $ are directly compared to the numerical simulations of SASI in Sect.~6. Before that, we use the simplified toy-model of \cite{sato09} to evaluate the nonlinear effect of the parasitic instabilities on the advective-acoustic cycle. The distortion of the SASI mode by growing parasites, illustrated by the bottom plots in Fig.~\ref{imageKH} and \ref{imageRT}, is expected to induce a decrease in the acoustic feedback and stabilize the advective-acoustic cycle responsible for the growth of SASI. 

The ``problem~1" studied by \cite{sato09} deals with the deceleration of an advected wave through an external potential, in a planar toy-model. This deceleration region of size $\Delta z_{\rm \nabla}$ generates an acoustic feedback, measured at a distance $z_{\rm meas}$ above it. The advected wave of amplitude $\epsilon_S$ is perturbed by a random noise acting as a seed for the parasitic instabilities. We choose $\Delta z_{\rm \nabla} = 0.4 $, $ M_{\rm 1} = 5 $, $ c_{\rm in}^{2}/c_{\rm out}^{2} = 0.75 $, $\omega\tau_{\rm aac}/2\pi = 2 $. The numerical technique based on a AUSMDV scheme is described in \cite{sato09}. 

The effects of the KHi and RTi on the acoustic feedback are studied together and separately by performing three sets of simulations: i) with the same mixture of entropy and vorticity as produced by a perturbed shock (Eq.~(9-13) of \cite{sato09}), ii) with the same entropy structure but no vorticity, iii) with the vorticity structure of (i) but no entropy. 

The pressure measured at $z_{\rm meas}=3$ is Fourier transformed in the $x$-direction and in time, 
in order to estimate the part of the acoustic feedback which is coherent with the initial advected wave. This coherent feedback is responsible for the closure the advective-acoustic cycle.

\begin{figure}
\centering
\includegraphics[width=\columnwidth]{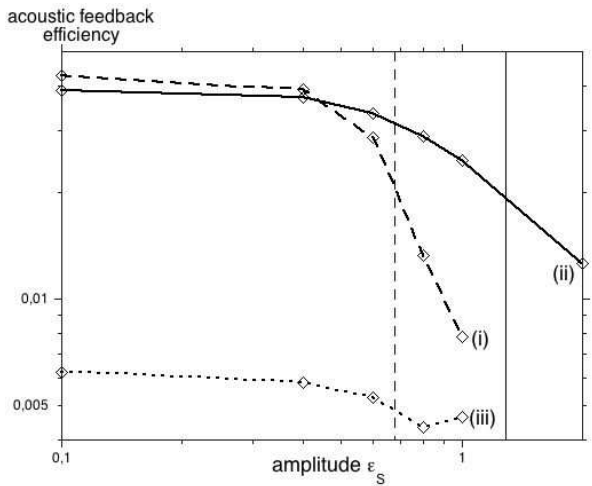}
 \caption{Efficiency of the acoustic feedback as a function of the amplitude of the advected wave. The dashed line corresponds to an entropy vorticity wave (case i), the full line to an entropy wave (case ii) and the dotted line to a vorticity wave (case iii). The two vertical lines represent the estimate of the amplitude cutoff where the corresponding efficiency has decreased by $50\%$: dashed for the entropy-vorticity wave, full for the entropy wave.}
             \label{toy_cutoff}
\end{figure}

\begin{figure}
\centering
\includegraphics[width=\columnwidth]{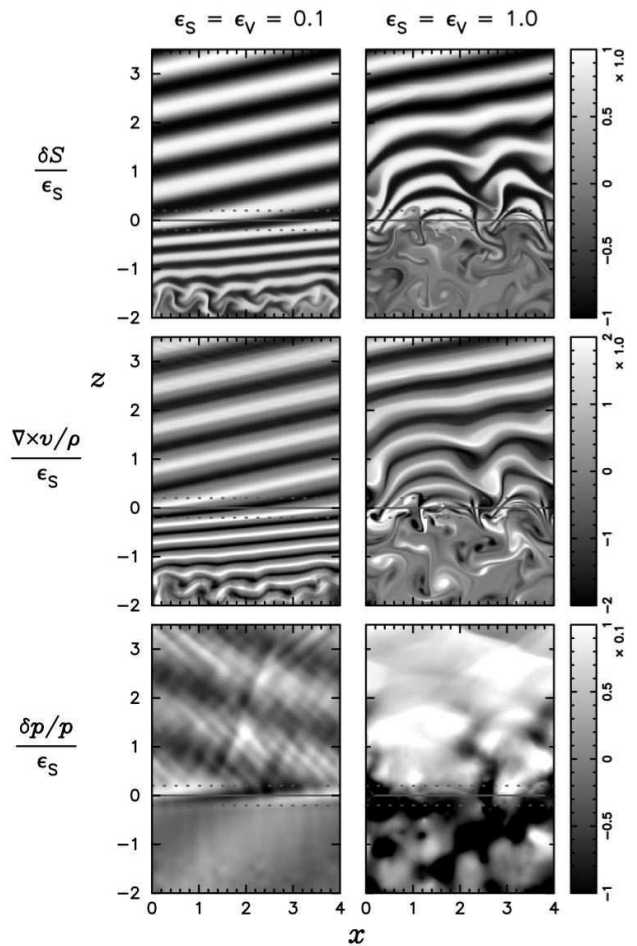}
 \caption{Acoustic feedback from an entropy-vorticity wave (case i) in the presence of the KHi parasites. The left column represents a wave of linear amplitude ($\epsilon_{S} =0.1 $, below the cutoff), while the right column represents a wave of non-linear amplitude ($\epsilon_{S}  = 1 $, above the cutoff). The three rows show entropy (upper), vorticity (middle), and pressure (bottom) perturbations. The horizontal dashed lines represent the extent of the potential jump. The KHi is able to grow only on the non-linear wave (right column).}
             \label{toyfigure_KH}%
\end{figure}

\begin{figure}
\centering
\includegraphics[width=\columnwidth]{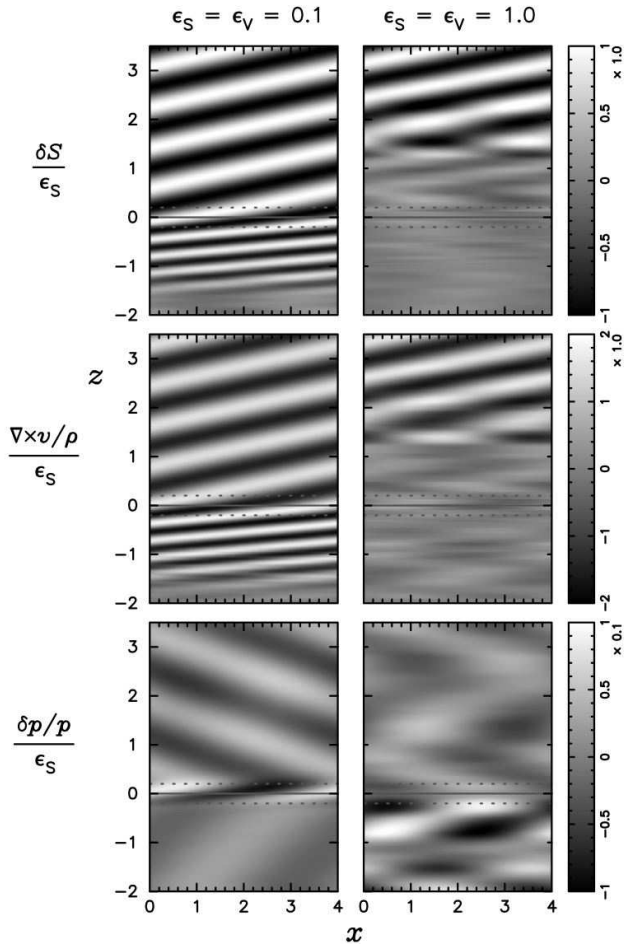}
 \caption{Same as Fig.~\ref{toyfigure_KH} but filtered to keep only the $n=1$ component of the horizontal structure. In the entropy and vorticity profile a clear decrease in amplitude is visible where the KHi has grown.  In the right column, the pressure perturbations are slightly smaller and the coherence of the vertical structure is lost.}
             \label{toyfigure_KH_filt}%
\end{figure}

\begin{figure}
\centering
\includegraphics[width=\columnwidth]{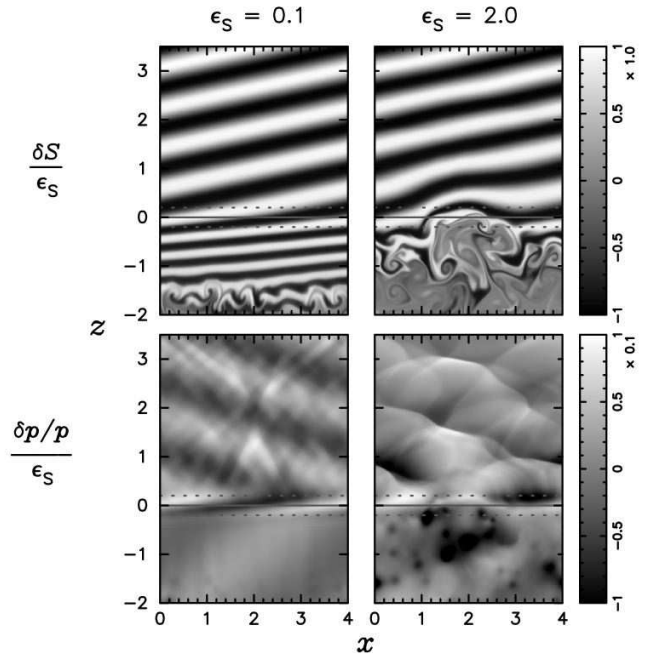}
 \caption{Acoustic feedback from an entropy wave (case ii) in the presence of RTi parasites. The left column represents a wave of linear amplitude ($\epsilon_{S}  =0.1 $, below the cutoff), while the right column represents a wave of non-linear amplitude ($\epsilon_{S}  = 2 $, above the cutoff). The two rows show entropy (upper) and pressure (bottom) perturbations. The horizontal dashed lines represent the extent of the potential jump. The RTi is able to grow only on the non-linear wave (right column).}
             \label{toyfigure_RT}%
\end{figure}

Above a certain amplitude threshold, the acoustic feedback efficiency decreases from the value predicted by the linear analysis to a small fraction of this value (Fig.~\ref{toy_cutoff}). This threshold is measured as the amplitude $\epsilon_{S}$ at which the acoustic feedback efficiency is $50\%$ of its linear value. We find a value of $\epsilon_{S} = 0.68 $ for case (i) and $ \epsilon_{S}  = 1.3 $ for case (ii).  The decrease of the feedback is due to the development of parasitic instabilities propagating against the flow (Figs. \ref{toyfigure_KH} and \ref{toyfigure_RT}). The growth of the parasites causes the advected wave to lose its coherence: the power remaining in its $n=1$ component is decreased significantly in the region where the parasites have grown (Fig.~\ref{toyfigure_KH_filt}). The vertical structure of the feedback is also distorted.  

In Fig.~\ref{toy_cutoff}, the small amplitude of the acoustic feedback in the case (iii) decreases by $30\%$ for $\epsilon_S=0.8$, and increases again for $\epsilon_S=1$: this increase is due to the pressure associated with the KHi, propagating against the flow.

\begin{figure}
\centering
\includegraphics[width=\columnwidth]{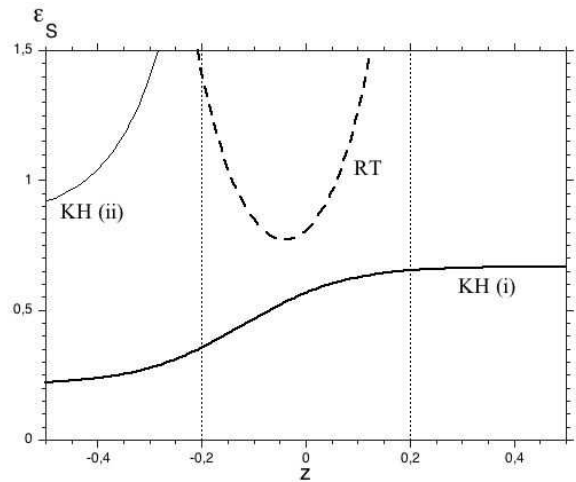}
 \caption{Amplitude $\epsilon_{S} $ corresponding to the marginal stability of secondary instabilities, as a function of $z$ in the problem 1 of \cite{sato09}. Full lines show the KHi marginal stability: the entropy-vorticity wave (thick line, case i),  the entropy wave (thin line, case ii). The dashed line shows the RTi marginal stability in both cases. The two vertical dotted lines show the vertical extent of the potential jump. }
             \label{localsaturation_toymodel}%
\end{figure}

Can the linear description of the KHi and RTi (Sect.~3 and 4) predict the value of the threshold? We use the analytical estimates of the KHi and RTi growth rates (Eqs.~(\ref{eqwiKH}) and (\ref{eqwiRT})) at marginal stability to estimate the threshold amplitude $\epsilon_S(z)$ above which the KHi or RTi can grow despite the stabilizing effect of advection. The $\epsilon_S$-threshold for neutral stability is expected to be a lower bound for the threshold measured in the simulations (Fig.~\ref{localsaturation_toymodel}).

The RTi can grow only in the region $ -0.2<z<0.2 $ where gravity is significant, represented by the two horizontal dashed lines in Figs.~\ref{toyfigure_KH} to \ref{toyfigure_RT}. The local amplitude threshold in Fig.~\ref{localsaturation_toymodel} displays a clear minimum around $ z \sim 0 $ at a value of $ \epsilon_{S}  = 0.78 $, which appears to be a good estimate of the amplitude above which the RTi can grow and damp the acoustic feedback.

In the case (ii) where the upstream advected wave contains no vorticity, the growth of the KHi is subdominant and does not affect the acoustic feedback because it takes place below the region of deceleration (Fig.~\ref{localsaturation_toymodel}). The RTi is thus the dominant instability. The linear threshold ($ \epsilon_{S}  = 0.78 $) deduced from Fig.~\ref{localsaturation_toymodel} is about $40\%$ smaller than the value of $\epsilon_{S}  = 1.3 $ measured in the simulations.

In the case (i), the local amplitude threshold of the KHi measured in Fig.~\ref{localsaturation_toymodel} is smaller than that of the RTi.
The KHi should thus be the dominant parasitic instability with a smaller amplitude threshold than in case (ii) 
($\epsilon_{S} \in\lbrack0.36,0.65\rbrack$). Indeed an instability develops upstream of the potential jump in case (i) with vortices which are typical of the KHi (Fig.~\ref{toyfigure_KH}). For comparison, vortices are less prominent in case (ii) and no disturbance is growing upstream of the shock,  consistent with the RTi (Fig.~\ref{toyfigure_RT}). As predicted the cutoff in case (i) is smaller than that of case (ii) ($0.68$ versus $1.3$). The linear threshold ($\sim0.5$) is roughly 25\% smaller than the simulated value of $ \epsilon_{S}  = 0.68 $.

As a summary, we find that the linear thresholds for marginal stability are consistently $ 25-40 \% $ smaller than the threshold at half efficiency measured in the simulations. Let us emphasize that in case (i), although the linear acoustic feedback is essentially generated by the entropy wave, vorticity plays an essential role in determining the saturation threshold through the KHi. This illustrates the non-trivial interplay of vorticity and entropy in the advective-acoustic cycle.

\section{Comparison with (more) realistic simulations: the effect of nuclear dissociation}

In order to test the parasitic scenario against results from numerical simulations, we apply the above estimates to the set up of \cite{fernandez09a}, where the energy loss at the shock ($\epsilon$ in their notations) due to the dissociation of iron is varied in a parameterized way from $\epsilon = 0$ (the setup of \cite{blondin06}) to $\epsilon = 0.25v_{\rm ff}^{2}$. Here $v_{\rm ff}$ is the free fall velocity at the shock. The shock radius is kept constant by adjusting the cooling function, which is varied by a factor up to $127$ (the shock radius and the adiabatic index are $r_{\rm sh}/r_{\rm star} = 2.5$ and  $\gamma = 4/3$). These simulations are a very good test for any saturation mechanism, because the saturation amplitude of SASI was found to be sensitive to the parameter $\epsilon$.

\begin{figure}
\centering
\includegraphics[width=\columnwidth]{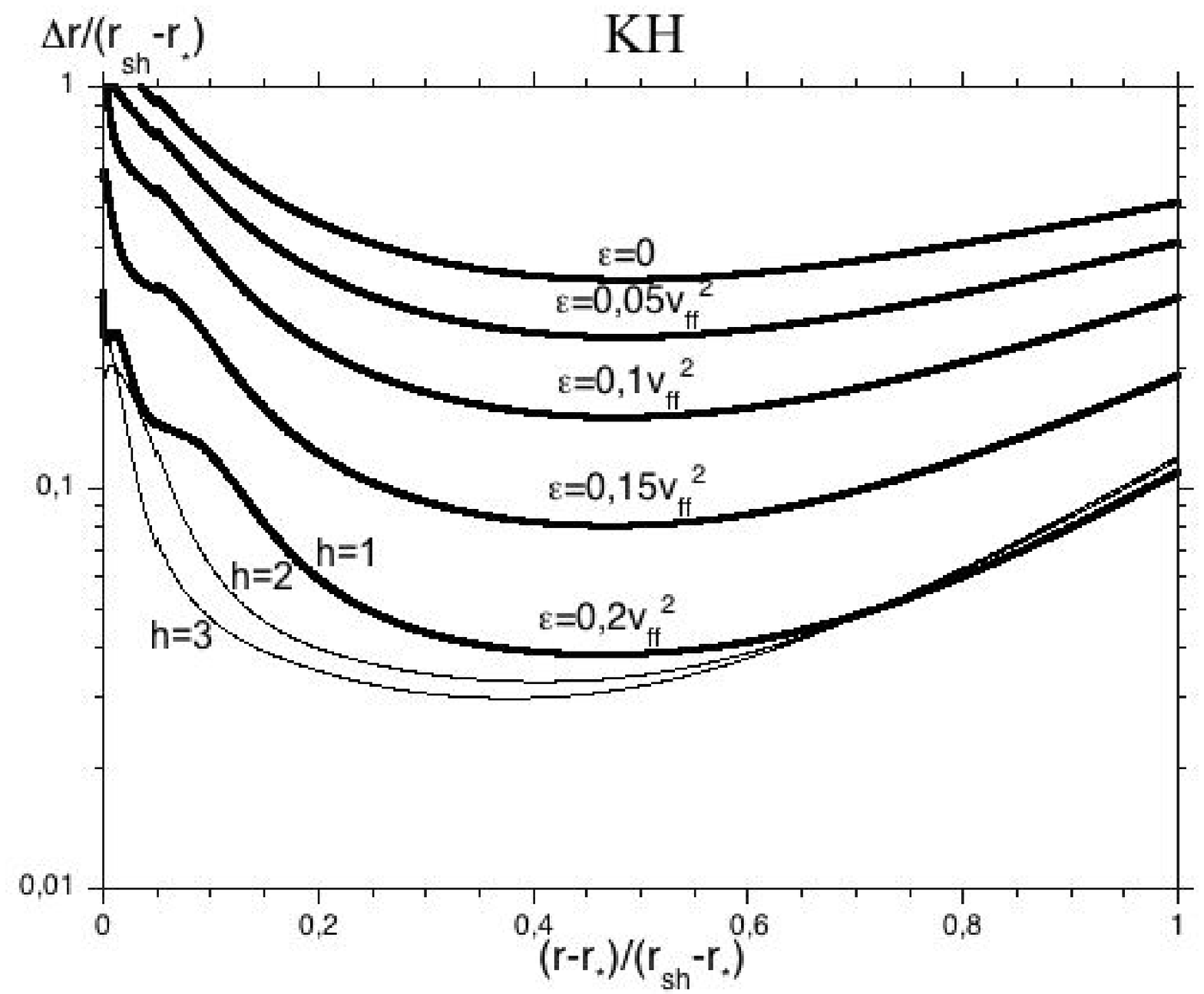}
\includegraphics[width=\columnwidth]{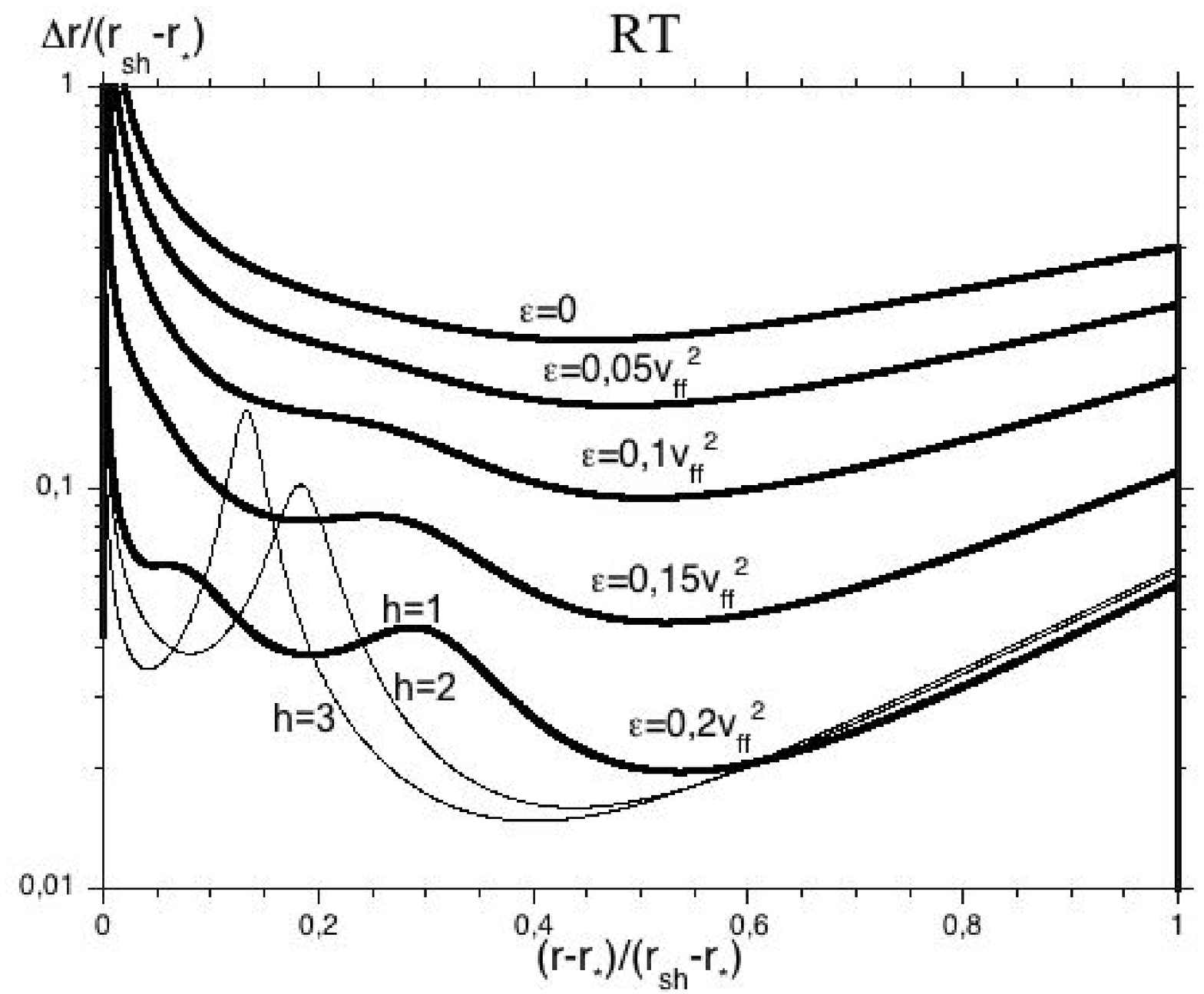}
 \caption{``Local saturation amplitude" as a function of the radius for the KHi (\emph{upper plot}) and RTi (\emph{bottom plot}) instabilities. The thick lines show the fundamental mode for different dissociation energies (upper curves correspond to lower energies). The thin lines show higher harmonics with the dissociation energy $\epsilon = 0.2v_{\rm ff}^{2}$.}
             \label{local_saturation}%
\end{figure}

By solving the radial structure of eigenmodes in the setup of \cite{fernandez09a}, we calculate the parameters $w_{\rm sasi}$ and $\nabla S_{\rm sasi}$ and use Eqs.~(\ref{satKH}) and (\ref{satRT}) to estimate the ``local saturation amplitude" of SASI oscillations above which the parasites grow faster than SASI at a given radius. As indicated by Fig.~\ref{local_saturation}, the saturation amplitude of SASI associated with each parasitic instability decreases strongly when $\epsilon$ increases.  

\begin{figure}
\centering
\includegraphics[width=\columnwidth]{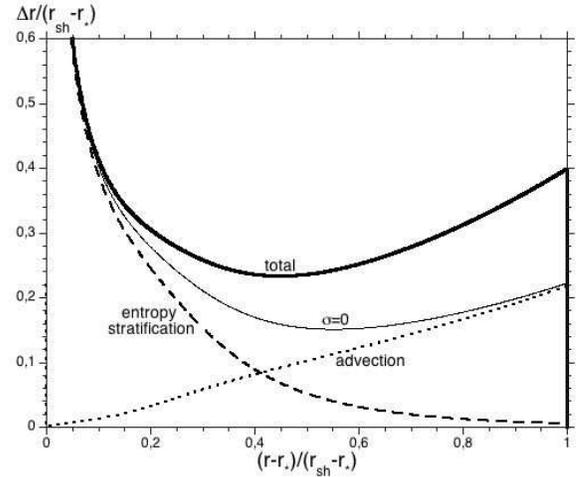}
 \caption{Roles of the advection, the entropy stratification, and the SASI growth rate in the ``local saturation amplitude". This is illustrated with the fundamental SASI mode at $\epsilon = 0$ and the RTi, but it is qualitatively the same if one considers other SASI modes, dissociation energies, or the KHi. The thick line shows the ``local saturation amplitude", the thin line the same quantity if one neglects the growth rate of SASI, the dotted and the dashed lines the contributions of the advection and the entropy stratification.}
             \label{local_saturationh1}%
\end{figure}

The global saturation amplitude can be estimated as the minimum of the local saturation amplitude, at least if this minimum is sufficiently broad and above the coupling radius. The curves in Fig.~\ref{local_saturation} show a minimum at an intermediate radius between the proto-neutron star  and the shock, approximately at $ r_{\rm mini} \sim (r_{\rm sh}+r_{\rm *})/2 $. This can be understood by the fact that higher up the shock parasites are efficiently stabilized by advection, while close to the proto-neutron star they are strongly stabilized by the entropy stratification (Fig.~\ref{local_saturationh1}). The most efficient growth of the parasites therefore takes place where neither advection nor stratification is strong. As the minima of the curves in Fig.~\ref{local_saturation} are quite flat, the parasites should be able to grow in a large region of the flow around the radius $ r_{\rm mini}  $ when SASI saturates.

\begin{figure}
\centering
\includegraphics[width=\columnwidth]{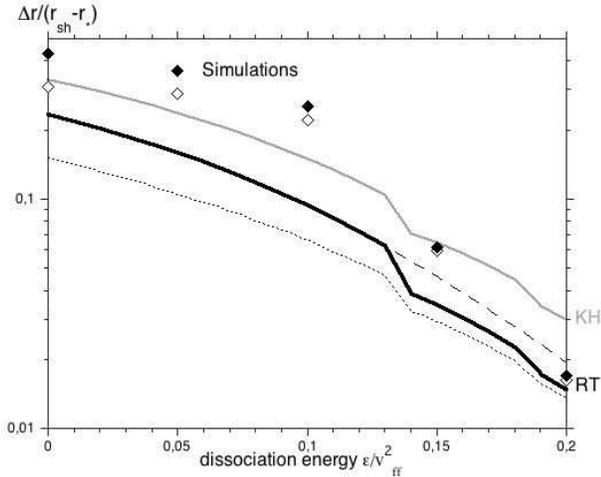}
 \caption{Saturation amplitude of SASI as a function of the dissociation energy $\epsilon$: Comparison of the simulations by \cite{fernandez09a} with the parasitic instabilities scenario. The black diamonds show the amplitude of the saturated $l=1$ mode in the simulations of \cite{fernandez09a} (rms fluctuation of the $l=1$ Legendre coefficient averaged over very long timescales, after the flow has settled to a quasi-steady state). The empty diamonds show the saturation amplitude normalized by $r_{sh} - r_{*}$, where $r_{sh}$ is the average shock radius during the nonlinear phase of SASI, instead of the shock radius in the stationary flow for the black diamonds (R.~Fernandez, private communication). The thick lines are the saturation amplitude predicted for the most unstable mode of the KHi (gray line), and the RTi (black). The dotted (respectively dashed) line shows the RTi saturation amplitude when $\sigma_{sasi}$ is set to 0 in Eqs.~(\ref{satRT}) (respectively when one considers only the fundamental mode of SASI).}
              \label{saturation_fernandez}%
\end{figure}

The saturation amplitude predicted by our analysis of the KHi and RTi is compared with the results of the simulations by \cite{fernandez09a} in Fig.~\ref{saturation_fernandez}. The RTi (thick full line) is expected to be the dominant parasite because it grows at smaller SASI amplitudes than the KHi. We note that some RTi structures are clearly visible in the simulations of 
\cite{fernandez09a} (online edition) for $\epsilon = 0.15v_{\rm ff}^{2} $ and $\epsilon = 0.2v_{\rm ff}^{2} $, in agreement with our conclusion that the RTi is the dominant secondary instability. However, these RTi structures are less obvious when $\epsilon=0$.

The amplitude in the simulations decreases by a factor $ \sim 25 $ between $\epsilon=0$ and $\epsilon = 0.2v_{\rm ff}^{2}$, while our estimate decreases by a factor 15. In addition to reproducing correctly the trend, the saturation amplitude given by this saturation mechanism is $15-50\%$ smaller than the simulated value for all $\epsilon$. This is consistent with Sect.~5, where we found that the stability threshold of the parasites was $\sim 25-40\%$ smaller than the amplitude at which their effect is important. Given the many approximations involved in our analytic description of the parasites, and the many other nonlinear effects we neglected, the comparison in Fig.~\ref{saturation_fernandez} is considered very encouraging. 

One of our assumptions is that the background stationary flow is unchanged, which is justified for low saturation amplitudes but is less justified if the saturation amplitude is very nonlinear. The uncertainty of our analytical estimate for large saturation amplitudes is illustrated in Fig. 16 by the empty diamonds, where the saturation amplitude is normalized using the averaged shock radius during the nonlinear phase of SASI instead of its value in the stationary flow. This new normalization brings the values from the simulations closer to the predicted ones (R.~Fernandez, private communication). 

What is the dominant effect causing the dramatic decrease of the SASI amplitude when $\epsilon$ is increased? Using Eq.~(\ref{satRT}) allows us to identify the contributions of the SASI growth rate $\sigma_{\rm sasi}$, the relative amplitude of the SASI entropy wave $\nabla S_{\rm sasi}$, the background entropy stratification $\nabla S_0$ and the background advection velocity $v_z$.
 
According to our linear stability analysis, the most unstable SASI mode is the fundamental one if $\epsilon$ is small, the first harmonics if $\epsilon > 0.15v_{\rm ff}^{2}$ and the second harmonics if $\epsilon > 0.19v_{\rm ff}^{2}$. The comparison between the dashed line and thick full line in Fig.~\ref{saturation_fernandez}, separated by about $ \sim 25\% $, shows that higher harmonics are more sensitive to the RTi than the fundamental mode (see also the thin lines in Fig.~\ref{local_saturation}). This effect on $\nabla S_{\rm sasi}$ is partly due to the fact that entropy gradients in a SASI mode increase with frequency. It is striking that the sharp drop in the SASI amplitude that is observed by \cite{fernandez09a}  coincides with  the shift from the fundamental to the first radial overtone. Our analysis suggests however that this shift may be the cause of only a small fraction of the decrease, and should not be considered as a general feature of the saturation of SASI with dissociation. Indeed, repeating our analysis for other aspect ratios of the shock to star radius indicates that this shift can also happen much earlier (e.g. for $r_{\rm sh}/r_{*} = 0.2$) or not at all (e.g. for $ r_{\rm sh}/r_{*} = 0.6 $).

The SASI amplitude required for marginal parasitic instability ($ \sigma_{\rm RT} = 0 $) can be compared to the saturation amplitude ($ \sigma_{\rm RT} = \sigma_{\rm sasi} $). From Fig.~\ref{saturation_fernandez} (dotted and thick lines), we estimate that the lower growth rate of SASI for high dissociation energy is responsible for a $ \sim 25\% $ decrease of the saturation amplitude of SASI.

An important consequence of energy losses at the shock is a slower post-shock advection speed $v_z$. As the RTi develops more easily in a slow flow, the saturation amplitude of SASI naturally decreases when dissociation is increased.
This major effect contributes to a factor $4.5$, evaluated by comparing the local saturation amplitude at the shock with and without dissociation (using $\sigma = 0$ in order to distinguish it from the effect of the change in the SASI growth rate).

An additional factor $2$ is due to the change in the flow profile, in particular to the decrease of the entropy stratification $\nabla S_0$ which favors the growth of parasitic instabilities.

\section{Discussion}

\subsection{SASI or neutrino-driven convection? }

\cite{fernandez09b} argued that the large amplitude $l=1$ oscillations appearing in their numerical simulations including 
iron dissociation and a heating function is due to neutrino-driven convection rather than SASI, since SASI is stabilized by iron dissociation according to \cite{fernandez09a}. Does iron dissociation at the shock really prevent SASI from growing to large amplitudes in a realistic core collapse? In realistic simulations the compression factor at the shock can reach $\sim 10 $, which corresponds to $ \epsilon = 0.14 $ in the present study and a SASI amplitude of $6\%$ of the shock distance $( r_{\rm sh} - r_{\rm *} ) $, quite smaller than without dissociation ($40\%$). We point out however that a significant fraction of this amplitude decrease may be an artifact of the parameterization,  which changed the cooling function by a factor 127 in order to keep the ratio $ r_{\rm sh}/r_{\rm *} $ constant. This parameterization has the great advantage of being insensitive to geometrical effects that may arise if the aspect ratio between the shock and the cooling surface is changed. However cooling is then artificially low when dissociation is taken into account without heating. As a consequence, the resulting flow profile may not be more realistic than the flow profile without dissociation. As our analysis suggests that entropy gradients play an important role in the saturation of SASI, we investigated the effect of keeping the cooling function constant 
when dissociation is varied, resulting in a change of the shock radius (from $2.5r_{*}$ to $1.46r_{*}$, for $0<\epsilon < 0.2v_{\rm ff}^{2}$). By performing the same analysis as in Section~6, we then find that the saturation amplitude of SASI should decrease significantly less than when the shock radius is kept constant : $\Delta r/( r_{\rm sh} - r_{\rm *} ) $ decreases by a factor $1.75$ only. Equivalently, $\Delta r/r_{\rm sh} $ and $\Delta r/r_{\rm *} $ are decreased by a factor $3.5$ and a factor $6$ respectively (to be compared with a decrease of 15 when $r_{\rm sh}$ is constant). Although the geometrical effects make a direct comparison difficult, the fact that all these numbers are significantly smaller than the former variation by a factor 15 suggests that the decrease of the cooling function, necessary to keep the shock radius constant, plays a key role in decreasing the saturation amplitude. More insight on this issue may be gained by including the effect of neutrino heating in a parameterized manner, such that dissociation could be varied while both the cooling function and the shock radius are constant. This calculation is left for a future study. 

The numerical simulations by  \cite{scheck08} suggest that SASI is able to grow to large amplitudes even in the presence of dissociation. These simulations are significantly more realistic than the set up studied here, since they include a realistic equation of state where dissociation is taken into account in a physical way, and a simplified treatment of neutrino heating and cooling. They also differed from those by \cite{fernandez09a} by their choice of a moving inner boundary mimicking the proto neutron star contraction. In some of the models of this article (e.g. W00), neutrino-driven convection was artificially suppressed but still SASI oscillations could grow to non negligible amplitudes. 
It is however difficult to determine which difference between the two setups affects most importantly the saturation amplitude of SASI.

Incidentally, it is worth noting that RTi mushrooms have been identified growing on the SASI entropy gradients in Fig.~7 of \cite{scheck08} and were interpreted in their Sect.~6.1 as secondary convection. Although in that article convection was not recognized as an agent of SASI saturation, the fact that the RTi appears at a SASI amplitude close to the saturation amplitude is consistent with a parasitic mechanism of saturation.

The interaction between SASI and neutrino-driven convection is complex and still poorly understood. Could neutrino-driven convection prevent the growth of SASI by breaking its mode structure ? One may argue that neutrino-driven convection does not feed upon the SASI mode energy, but rather converts free energy from the stationary gradients into vorticity. Could neutrino-driven convection feed SASI, either by creating vorticity which would enter the advective-acoustic cycle, or by creating sound waves \citep{fernandez09b}? These difficult questions are beyond the scope of our study.

\subsection{Distinguishing RTi from KHi in the simulations}

The RTi is often characterized by finger-like or mushroom-like structures as in Fig.~\ref{imageRT}, while the KHi is characterized by vortices as in Fig.~\ref{imageKH} and \ref{toyfigure_KH}. However, in a complex flow containing both entropy gradients, shear and advection, RTi mushrooms may look like vortices (Fig.~\ref{toyfigure_RT}). 

The following criterion may be more useful: the RTi should occur preferentially where the entropy perturbation is maximum, while the KHi occurs where the shear is maximum. In a sloshing mode these two maxima are very distinct: the entropy oscillation is maximum at the pole (where the shock speed is maximum), while vorticity is maximum at the equator (where the inclination of the shock is maximum). The parasitic structures visible in the simulations by \cite{scheck08} and in the movies published online by \cite{fernandez09a} are more vigorous near the pole, in agreement with our analysis.

Furthermore, the RTi structure should grow preferentially on the half wavelength of a SASI mode where the entropy gradient is negative. By contrast, the KHi should grow on the whole extent of the SASI wavelength. An inspection of Fig.~7 of \cite{scheck08} and of the movies by \cite{fernandez09a} confirms this distinct feature of the RTi.

\subsection{Numerical resolution needed to resolve the parasites}

An interesting concern raised by this saturation mechanism is that simulations should be able to resolve the parasitic instabilities properly in order to give reliable results on the nonlinear behavior of SASI. The RTi is a short wavelength instability, but as is shown in Sect.~4.3 advection tends to stabilize the small scales and makes the RTi dominated by large scales. Entropy stratification on the other hand favors small scales. As in the set up studied here the RTi is found to develop where both stabilizing effects are important, it is hard to make any prediction for its dominant wavelength. Any convergence study should verify that the grid size allows for the growth of parasites.

As an example, \cite{scheck08} witnessed the growth of Rayleigh-Taylor mushrooms with a typical angular scale of $ l \sim 20-30 $. They were able to capture this small angular scale by using $360$ angular zones for $180\,^{\circ} $. Most 2D simulations use a resolution with $>100-200$ angular zones, and would probably resolve these scales (200 zones in \cite{murphy08b}, 121 in \cite{burrows06}, 128-192 in \cite{marek09}, and 60-120 zones in \cite{ohnishi06}). However 3D simulations may not be able to resolve such small scales. For example \cite{iwakami08, iwakami09} mostly use a resolution of $30$ angular zones for $180\,^{\circ}$, which may be too coarse to capture such a small scale behavior. We note that the saturation amplitude of the low-$l$ modes ($l=1-3$) in  Fig.~16 of \cite{iwakami08} is slightly smaller at ``high resolution" (60 zones) than at ``low resolution" (30 zones).  While this may be explained by a suppression of parasitic instabilities at low resolution, we cannot exclude that a different saturation process may take place in 3D, as discussed below.

\subsection{Effects of other physical ingredients}

The saturation mechanism described in this paper can be used to anticipate the effects of many other physical ingredients of the core collapse model (e.g. 3D versus 2D, the rotation rate, the magnetic field) although a detailed analysis is beyond the scope of this paper.

\begin{itemize}
\item \emph{3D versus 2D}: 
3D simulations allow for non-axisymmetric modes that are artificially forbidden in axisymmetric simulations, thus a greater number of modes is available to the linear development of SASI. A single mode $l=1$, $m=0$ often dominates in 2D, whereas $3$ modes $l=1$, $m=0,\pm1$ have the same growth rate in 3D if the collapsing core does not rotate \citep{foglizzo07}. Besides, \cite{iwakami08} found that the saturated mode amplitude is independent of $m$. We cannot exclude that nonlinear processes associated with the coupling between different mode, ignored in our analysis, are more important in 3D than in 2D. Contrary to \cite{iwakami08} however, \cite{blondin07a} found that \emph{one} spiral mode dominates the 3D dynamics.

Assuming the parasitic growth of instabilities is the dominant saturation mechanism, our analysis based on a linear description of the parasites would predict the same saturation amplitudes of SASI in 2D or 3D. However the nonlinear behavior of the RTi is known to differ in 3D and 2D (e.g. \cite{goncharov02}, \cite{cabot06}), and this may affect the saturation of SASI. \cite{iwakami08} reported a  smaller saturation amplitude of each individual SASI mode in 3D as compared to 2D, although the numerical convergence of this result should be further checked (Section 7.3). If confirmed, it would raise the following questions : is this difference in amplitude a consequence of the different non linear Rayleigh-Taylor behavior in 3D? Or is this the signature of  a different saturation mechanism based on the interaction of $m\ne 0$ modes ? A more systematic parametric study, similar to \cite{fernandez09a} but in 3D, could help check the relevance of parasitic instabilities in 3D.

\item \emph{Rotation rate}: \cite{yamasaki08} have shown that rotation increases the growth rate of the spiral modes rotating in the same direction as the steady flow, while stabilizing the counter-rotating ones. If the rotation is strong enough, a single spiral mode dominates the evolution of SASI \citep{blondin07a, iwakami08}. According to our analysis (Eq.~(\ref{satRT})), the larger growth rate of the spiral mode could lead to a larger saturation amplitude of SASI. Nevertheless, a detailed calculation using the exact entropy and vorticity profiles of the SASI eigenmodes in a rotating flow is required in order to make an accurate prediction.

\item \emph{Magnetic field strength}: 
The effect of the magnetic field on the linear phase of SASI is yet to be understood \citep{guilet10}, but its effect on parasitic instabilities can already be anticipated from the point of view of the magnetic tension which tends to prevent motions that distort the magnetic field lines. This effect is stabilizing for the perturbations with a wave vector parallel to the magnetic field, but does not affect those whose wave vector is perpendicular.  One would then expect that  the magnetic field does not change the maximum RTi growth rate, but selects RTi modes with a wave vector perpendicular to the field lines. In contrast, the KHi can be suppressed if the magnetic field along the direction of the transverse velocity is strong enough. In a situation where the KHi were the dominant parasitic instability, a magnetic field could potentially allow for a larger saturation amplitude.
\end{itemize}

\section{Summary}

 In this article we have developed for the first time a predictive mechanism for the saturation of SASI. In this scenario the saturation happens when a parasitic instability is able to grow fast enough to compete with SASI. Two types of instabilities are of potential importance: the RTi growing on the entropy gradients created by SASI, and the KHi growing on the vorticity involved in the SASI mode. For each of these parasites, two stabilizing effects were found to be crucial: the entropy stratification in the stationary flow and the advection of matter toward the neutron star. An estimate of the growth rates taking into account these effects has been obtained in Sect.~3 and 4 for the KHi and RTi respectively. The saturation amplitude of a given SASI mode has been evaluated by comparing its growth rate with that of the parasitic instabilities.

Using numerical simulations, we studied the effect of parasitic instabilities on the acoustic feedback in  the simplified context of the toy model introduced by \cite{foglizzo09} and \cite{sato09}. This confirmed the idea of a threshold in amplitude above which the acoustic feedback is reduced, that is determined by the ability of parasitic instabilities to grow despite advection. A reasonable estimate of this threshold has been obtained by measuring the threshold of marginal stability, which is found to be $25-40\%$ lower than the amplitude at which the feedback is decreased by $50\%$.

The saturation mechanism by parasitic instabilities can reproduce the decrease of the SASI power with dissociation energy observed in the simulations of \cite{fernandez09a}. Our amplitude estimate based on linear growth rates remains $15-50\%$ lower than the saturation amplitude observed in their simulations, which is consistent with our simulation of the acoustic feedback in the toy-model. Furthermore our analysis suggests that the RTi is the dominant secondary instability. This is consistent with the presence of RTi mushrooms in the simulations of \cite{fernandez09a} as well as \cite{scheck08}.

The strong decrease of the SASI power, when both the dissociation energy is increased and the cooling is decreased, can be traced back to $4$ different effects that help the parasitic growth of the RTi: (i) the slower the advection velocity in the postshock flow, the faster the propagation of the RTi against the flow, (ii) the softer the negative entropy profile in the background flow, the easier its destabilization by the SASI entropy wave, (iii) the slower the growth of SASI, the lower the threshold for competing parasites, (iv) the steeper the entropy wave in the SASI mode, the faster the RTi. The first two effects are the dominant ones.
 We point out that other studies \citep{scheck08, ohnishi06}, which included the effect of nuclear dissociation, witnessed powerful SASI oscillations, contrary to \cite{fernandez09a}.  These studies differed from that of \cite{fernandez09a} by the use of a realistic equation of state, the inclusion of some heating, and in the case of \cite{scheck08} the contraction of the inner boundary. The dominant cause of the differing simulations has not been identified yet. Further investigations could shed more light on this question by applying the analysis described in this article to set ups that include one or several of these additional ingredients (heating, realistic EOS, contraction of the innner boundary).

Although the saturation mechanism proposed in this paper compares favorably with the results of \cite{fernandez09a}, it should be tested with results from other setups in order to confirm its validity, in particular in the case of 3D simulations. We propose that future simulations should look for signs of the parasitic instabilities, and check that the angular resolution is sufficient to resolve them. If confirmed, our results would open new perspectives for anticipating the effect on the SASI amplitude of other physical ingredients such as the equation of state, the heating rate, the rotation, and magnetic field of the progenitor star. They could also be useful as an input for analytical models studying the possible consequences of SASI, such as the model for gravitational wave emission proposed by \cite{murphy09} (see also \cite{marek09b}).

\acknowledgments{ The authors are grateful to S.~Fromang for stimulating discussions and his help with the code RAMSES. Interesting discussions with J.~Griffond and B.-J.~Grea are also acknowledged. R.~Fernandez is thanked for sharing data from his simulations, as well as for insightful comments on an early draft. The anonymous referee is also thanked for his remarks that signiÞcantly improved this article. This work has been partially funded by the Vortexplosion project ANR-06-JCJC-0119. }

\end{document}